\documentclass[aps,pre,preprint,numerical,a4paper,superscriptaddress,includegraphics,nofootinbib]{revtex4-1}
\usepackage{lipsum}
\usepackage{chngcntr}
\usepackage{graphicx}
\usepackage{parskip}
\usepackage{soul}
\usepackage{xcolor}
\usepackage{rotating}
\usepackage{wrapfig}
\usepackage{graphicx}
\usepackage{amsmath}
\usepackage{bbm}
\usepackage{amsfonts}
\usepackage{amssymb}
\usepackage{amsbsy}
\usepackage[T1]{fontenc}
\usepackage{theorem}
\usepackage{stmaryrd}
\usepackage{tipa}
\usepackage{textcomp}
\usepackage{siunitx}
\usepackage{subfigure}
\usepackage{epsfig}
\usepackage{lmodern}
\usepackage{framed}
\usepackage[subfigure]{tocloft}
\usepackage{subeqnarray}
\usepackage[refpage,cfg,prefix]{nomencl}
\usepackage{multirow}
\usepackage{xifthen}
\usepackage{enumerate}
\usepackage{enumitem}
\usepackage{color}
\usepackage{wasysym}
\usepackage{bibentry}
\usepackage{lipsum}
\usepackage{relsize}
\usepackage{verbatim}
\usepackage{hyperref}
\hypersetup{
colorlinks=true,
urlcolor=blue,
linkcolor=black,
citecolor=black,
breaklinks=true
}
\usepackage{breakurl}
\usepackage{url}
\usepackage{setspace}

\usepackage{tabularx}
\usepackage{booktabs}
\usepackage{subfiles}
\usepackage{mathtools}
\usepackage[UKenglish]{babel}
\usepackage[UKenglish]{isodate}
\usepackage{appendix}
\usepackage{lastpage}
\usepackage{fancyhdr}

\newcommand{\D}[2]{\frac{\partial #1}{\partial #2}}

\newcommand{\listofalgorithms}{\textbf{\Huge{List of Algorithms}}}
\newlistof{Algorithm}{exp}{\listofalgorithms}
\newcounter{instructioncounter}

\newcommand\numberthis{\addtocounter{equation}{1}\tag{\theequation}}

\allowdisplaybreaks

\newcommand*{\blankpage}{%
\vspace*{\fill}
{\fontfamily{phv}\selectfont{\centering This page intentionally left blank.\par}}
\vspace{\fill}}
\makeatletter
\renewcommand*{\cleardoublepage}{\clearpage\if@twoside \ifodd\c@page\else
\blankpage
\thispagestyle{empty}
\newpage
\if@twocolumn\hbox{}\newpage\fi\fi\fi}
\makeatother

\begin{document}

\title{Modelling adhesion in stochastic and mean-field models of cell migration}

\author{Shahzeb Raja Noureen}
\email[Corresponding author: ]{srn32@bath.ac.uk}
\affiliation{Centre for Mathematical Biology, University of Bath, Claverton Down, Bath, BA2 7AY, UK.}

\author{Richard L. Mort}
\affiliation{Division of Biomedical and Life Sciences, Faculty of Health and Medicine, Furness Building, Lancaster University, Bailrigg, Lancaster, LA1 4YG, UK.}

\author{Christian A. Yates}
\affiliation{Centre for Mathematical Biology, University of Bath, Claverton Down, Bath, BA2 7AY, UK.}

\date{\today}

\begin{abstract}

Adhesion between cells plays an important role in many biological processes such as tissue morphogenesis and homeostasis, wound healing and cancer cell metastasis. From a mathematical perspective, adhesion between multiple cell types has been previously analysed using discrete and continuum models including the Cellular Potts models and partial differential equations (PDEs). While these models can represent certain biological situations well, Cellular Potts models can be computationally expensive and continuum models only capture the macroscopic behaviour of a population of cells, ignoring stochasticity and the discrete nature of cell dynamics. Cellular automaton models allow us to address these problems and can be used for a wide variety of biological systems. In this paper, we consider a cellular automaton approach and develop an on-lattice agent-based model (ABM) for cell migration and adhesion in a population composed of two cell types. By deriving and comparing the corresponding PDEs to the ABM, we demonstrate that cell aggregation and cell sorting are not possible in the PDE model. Therefore, we propose a set of stochastic mean equations  (SMEs) which better capture the behaviour of the ABM in one and two dimensions.

\end{abstract}

\maketitle

\section{Introduction}

Collective migration of cells is an important biological process responsible for the correct formation of tissues during development and their maintenance throughout life. Although each cell is an individual unit, its movement cannot always be considered independent of other cells in its environment. Cells have been observed to exhibit a wide range of interactions with other cells in their neighbourhood. Some of the well-known interactions include volume exclusion (cells prevent other cells from occupying the same space) \citep{loeb1921amt,abercrombie1954osb,abercrombie1979cim,abercrombie1997cl1}, contact inhibition of locomotion (when two cells collide they change directions or stop moving) \citep{lee1995cam,carmona-fontaine2008cil,deutsch2021bca}, pushing (a moving cell can push cells in front) \citep{yates2015ipe}, pulling (a moving cell can pull neighbouring cells with it) \citep{yamanaka2014vas,davis1981scc,chappelle2019pmc}, and cell-cell swapping (two neighbouring cells can swap positions with each other) \citep{rajanoureen2023slc,osborne2017cia}. Another interaction to attract significant biological and mathematical interest is cell-cell adhesion \citep{foty2005dah,schmidt2010icm} in which cells tend to `stick' to one another through surface binding proteins known as cell adhesion molecules (CAMs) \citep{armstrong2006cam, foty2005dah}. Cell-cell adhesion is implicated in a wide range of important biological processes including the formation of tissues during development and their stability post-development; the invasion and metastasis of tumour cells; and developmental processes such as gastrulation and vasculogenesis \citep{armstrong2006cam, foty2005dah}.

Several mathematical models have been successfully employed to represent cell-cell adhesion. Discrete lattice-based stochastic models and deterministic continuum models based on partial differential equations (PDEs) are both commonly used paradigms. \citet{armstrong2006cam} propose a continuum description with non-local adhesive interactions to model differential adhesion, representing the average behaviour of two well-mixed populations of cells. Differential adhesion describes the adhesion of two or more populations of cells each of which has different adhesion properties. The differential adhesion hypothesis (DAH) views the two populations of cells as immiscible fluids with different surface tensions. It states that when two populations of cells (such as embryonic cells) with different adhesion properties are mixed, the energetically favourable configuration of the system in the steady state depends on the relative adhesion strengths of the cells \citep{foty2005dah,foty2004cca,steinberg1963rtd,armstrong2006cam}. \citet{armstrong2006cam} use their model to replicate experimentally observed cell sorting patterns in single-species and two-species scenarios as predicted by the DAH \citep{foty2005dah,foty2004cca,steinberg1963rtd,schotz2008qdt,armstrong2006cam}. One limitation of this continuum model is that it relies upon characterising the mathematical functions governing the adhesive forces acting on cells, which can be difficult to discern.

\citet{anguige2008omc} develop a continuum model with local adhesive interactions. Their approach considers cells on a one-dimensional discrete lattice in which each lattice site can be occupied by multiple cells at the same time and cells adhere to one another with strength, $\alpha>0$. They find that for $\alpha<0.75$ cells tend to diffuse out towards a uniform steady state as time progresses and for $\alpha>0.75$ aggregations of cells in space can be observed.

\citet{graner1992sbc} and \citet{glazier1993sda} use an extended Cellular Potts model -- a discrete approach in which each cell is represented by multiple lattice sites, allowing the cell to change shape according to a given energy minimisation function \citep{graner1992sbc,glazier1993sda,osborne2017cia} -- to replicate cell sporting patterns in biology from a uniformly mixed initial configuration of two populations. Cellular Potts models can incorporate mechanical features such as membrane tension and cell-cell and cell-substrate adhesion which, in conjunction with the shape-changing property of the cells, adds an extra layer of realism to these models \citep{osborne2017cia}. For large cell populations, simulating these models can be computationally expensive.

\citet{simpson2010mbc} use a cellular automaton model to represent the adhesion of breast cancer cells. We will review this model in detail here since the two-species adhesion model we propose later in this paper is based on the single-species model of \citet{simpson2010mbc}.

Modelling a monolayer of breast cancer cells, the authors consider agents on a regular two-dimensional lattice with $L_x$ compartments in the horizontal direction and $L_y$ compartments in the vertical direction. Each lattice site is square with length $\Delta$ such that a site can only contain one agent at a time making the model a volume exclusion process. An agent occupying site $(i,j)$ can attempt to move to a neighbouring site (up, down, left or right) with movement rate $m$ such that $m \delta t$ is the probability of the agent attempting a move during the infinitesimally small time window of duration $\delta t$. If the neighbouring site chosen to move into is empty, the agent successfully moves and if the neighbouring site is already occupied by another agent the move is aborted due to volume exclusion. Movement is further restricted by adhesion to neighbours in the Von-Neumann neighbourhood. The strength of the adhesive bond that two neighbouring agents share is denoted as $q \in [0,1]$ with $q=0$ meaning no adhesion and $q=1$ meaning maximal adhesion (i.e. no movement is possible). Consequently, the probability of movement depends on the agent breaking current bonds with its neighbours. Once selected to move into an empty site, the probability of overcoming the adhesive forces to complete the move is $(1-q)^{\Sigma_{z \in Z_{ij}} C_z}$ where $Z_{ij}= \{(i-1,j), (i+1,j), (i,j-1), (i,j+1) \}$ is the set of the indices corresponding to the four neighbouring sites of the site $(i,j)$ and $C_z$ is a binary indicator whereby $C_z=1$ if the neighbouring site $z \in Z_{ij}$ is occupied or $C_z=0$ otherwise.

To derive the continuum model, let $\hat C_{ij}(t)$ be the continuous approximation  of the density at the site $(i,j)$ at time $t$. We can write down the master equation for the  $\hat C_{ij}(t+\delta t)$ in terms of the density of the neighbouring sites at time $t$,

\begin{align}\label{eqn:ss_master_eqn}
\hat C_{ij}(t+\delta t) &=  \hat C_{ij}(t) +\frac{m}{4}\delta t  \hat C_{i-1,j}(t) (1- \hat C_{ij}(t) )(1-q)^{\Sigma_{z \in Z_{i-1,j}} \hat C_{z}} \nonumber \\
    &\quad+ \frac{m}{4}\delta t  \hat C_{i+1,j}(t) (1- \hat C_{ij}(t) )(1-q)^{\Sigma_{z \in Z_{i+1,j}} \hat C_{z}} \nonumber \\
    &\quad+ \frac{m}{4}\delta t  \hat C_{i,j-1}(t) (1- \hat C_{ij}(t) )(1-q)^{\Sigma_{z \in Z_{i,j-1}} \hat C_{z}} \nonumber \\
    &\quad+ \frac{m}{4}\delta t  \hat C_{i,j+1}(t) (1- \hat C_{ij}(t) )(1-q)^{\Sigma_{z \in Z_{i,j+1}} \hat C_{z}} \nonumber \\
    &\quad- \frac{m}{4}\delta t  \hat C_{ij}(t) (1- \hat C_{i-1,j}(t) )(1-q)^{\Sigma_{z \in Z_{i,j}} \hat C_{z}} \nonumber \\
    &\quad- \frac{m}{4}\delta t  \hat C_{ij}(t) (1- \hat C_{i+1,j}(t) )(1-q)^{\Sigma_{z \in Z_{i,j}} \hat C_{z}} \nonumber \\
    &\quad- \frac{m}{4}\delta t  \hat C_{ij}(t) (1- \hat C_{i,j-1}(t) )(1-q)^{\Sigma_{z \in Z_{i,j}} \hat C_{z}} \nonumber \\
    &\quad- \frac{m}{4}\delta t  \hat C_{ij}(t) (1- \hat C_{i,j+1}(t) )(1-q)^{\Sigma_{z \in Z_{i,j}} \hat C_{z}}
\end{align}

The RHS of Equation \eqref{eqn:ss_master_eqn} corresponds to the possible ways in which the occupancy of the site $(i,j)$ can change. Either:

\begin{enumerate}
\item nothing happens and the occupancy at time $t+\delta t$ is equal to the occupancy at time $t$ (first term on the RHS), or;

\item the possibility that the site $(i,j)$ is empty and an agent from a neighbouring site moves into the site $(i,j)$ breaking its existing bonds with neighbouring cells (if any) in doing so (second term on the RHS of line 1 and the terms in lines 2 to 4), or;

\item the possibility that the site $(i,j)$ is occupied at time $t$ and the occupying agent moves out of site $(i,j)$ to one of the empty neighbouring sites, breaking its existing bonds with neighbouring agents in doing so (lines 5 to 8 on the RHS).
\end{enumerate}

Taylor expanding the terms in Equation \eqref{eqn:ss_master_eqn} around site $(i,j)$ keeping terms up to second order and taking the limit as $\Delta, \delta t \to 0$ gives us the PDE for the mean continuum density $C(t,x,y)$,

\begin{equation}{\label{eqn:ss_pde}}
    \D C t (t,x,y) = D_0 \nabla \cdot [D(C(t,x,y))\nabla C(t,x,y)],
\end{equation}

where,

\begin{equation}
D_0=\lim_{\Delta, \delta t \to 0} \frac{m \Delta^2}{4\delta t},
\end{equation}

is the diffusion constant given that $\Delta^2/\delta t$ is held constant in the diffusive limit and,

\begin{equation}\label{eqn:diff_coeff_ss}
D(C)=(1-q)^{4C}[1-4C(C-1)\ln(1-q)],
\end{equation}

is the density-dependent diffusion coefficient.

\begin{figure}[t!]
\begin{center}

\subfigure[]{
\includegraphics[width=0.4\textwidth]{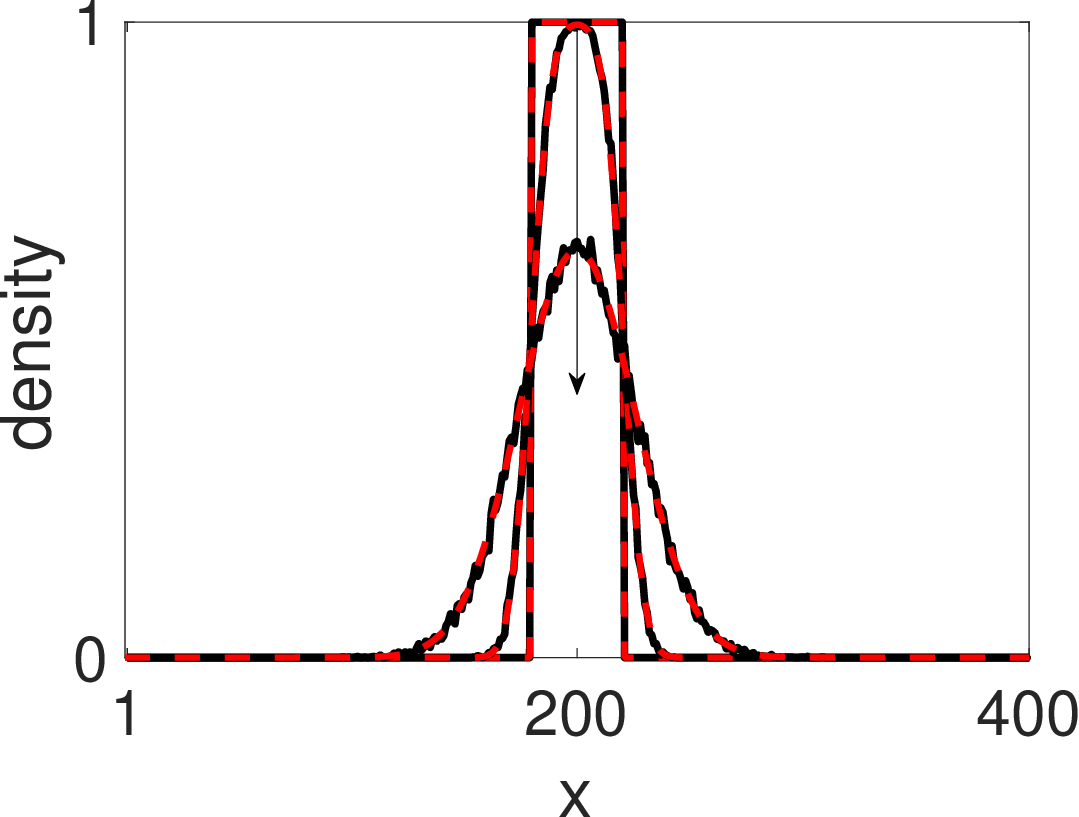}
\label{fig:ss_q_0}
}
\subfigure[]{
\includegraphics[width=0.4\textwidth]{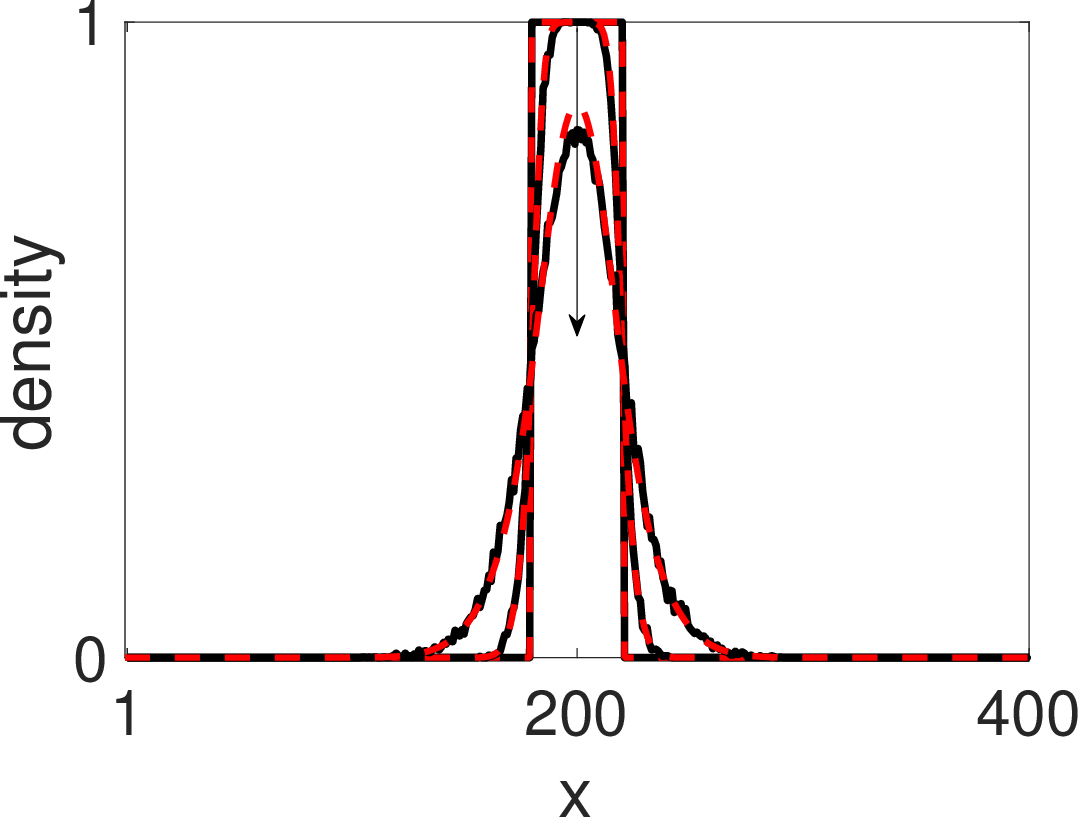}
\label{fig:ss_q_0.25}
}

\subfigure[]{
\includegraphics[width=0.4\textwidth]{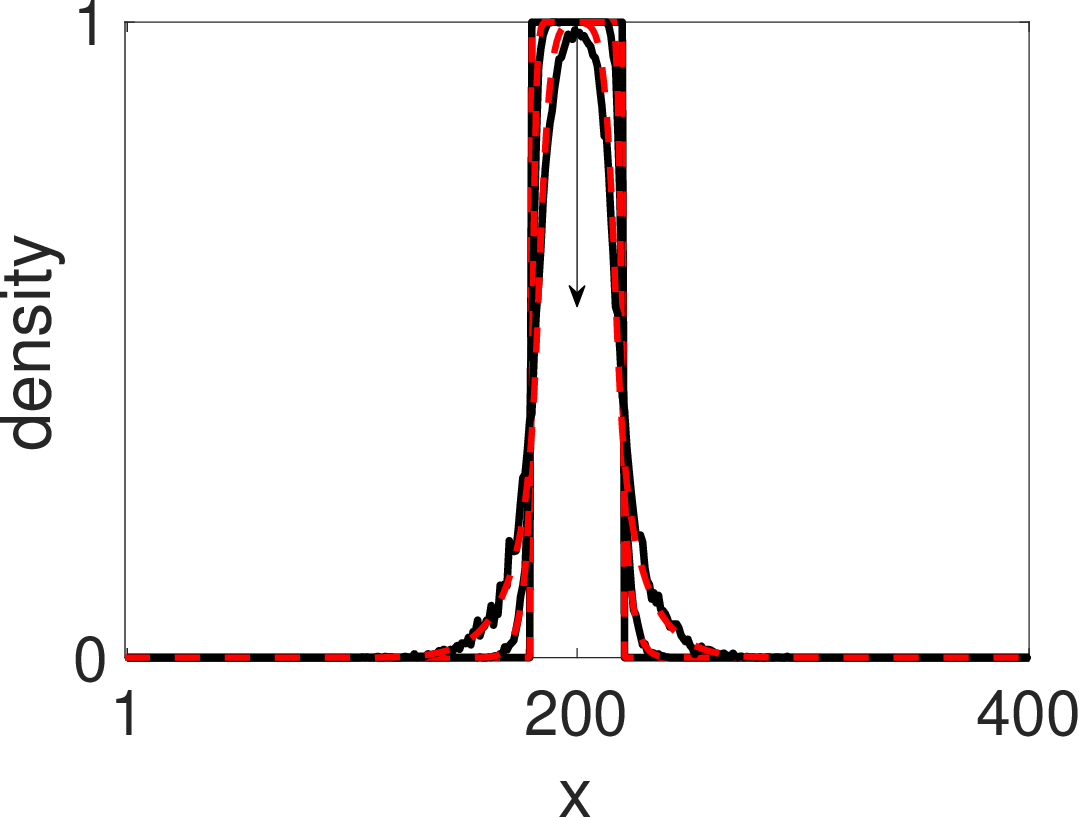}
\label{fig:ss_q_0.5}
}
\subfigure[]{
\includegraphics[width=0.4\textwidth]{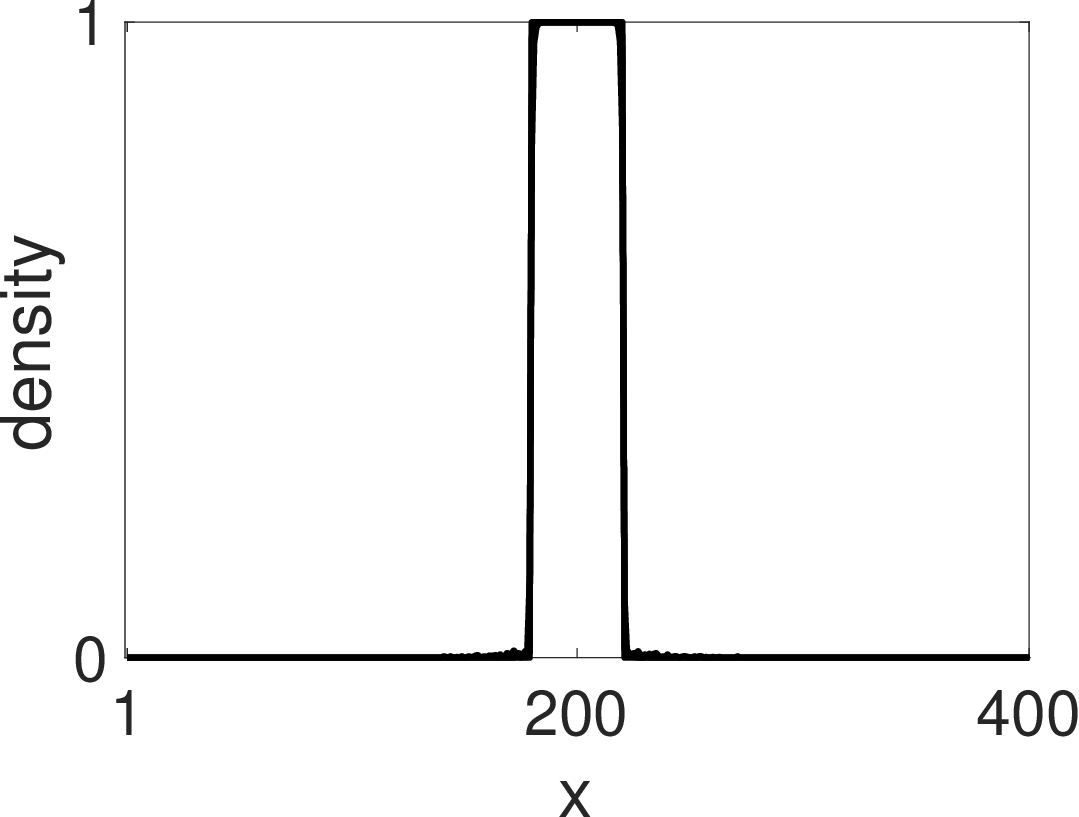}
\label{fig:ss_q_0.9}
}
\end{center}
\caption{Comparison between the column-averaged ABM (black) and the vertically-averaged solution of PDE \eqref{eqn:ss_pde} (red) for $q=0, 0.25, 0.5, 0.9$ at times $t=$ 0, 100, 1,000 with zero-flux boundary conditions. The movement rate $m=1$ in all cases. The domain is initialised with $L_x=400$ and $L_y=20$. Agents are initialised on the domain such that all the sites between $181 \leqslant x \leqslant 220$ are occupied by an agent. The agents evolve according to the ABM with zero-flux boundary conditions and column-averaged density of agents at $t=$ 0, 100 and 1,000 are shown in black. The black arrows indicate the direction of increasing time. Note: in the $q=0.9$ case the PDE became unstable hence the numerical solution is absent from the panel \subref{fig:ss_q_0.9}.}
\label{fig:ss_abm_pde_comparison}
\end{figure}

We compare the column-averaged density of cells in the ABM,

\begin{equation}
\bar C_i(t)=\frac{1}{L_y}\sum_{j=1}^{L_y} \langle C_{ij}(t) \rangle,
\end{equation}

where $\langle C_{ij}(t)\rangle$ is the average over $K$ repeats of the ABM, to the vertically-averaged solution of Equation \eqref{eqn:ss_pde} given by,

\begin{equation}\label{eqn:vert_ave}
    \bar C(t,x)=\frac{1}{L_y}\int_{y=0}^{y=L_y} C(t,x,y).
\end{equation}

We initialise the domain with $L_x=400$ and $L_y=20$ with zero-flux boundaries such that the middle 10\% of the lattice, corresponding to $181 \leqslant i \leqslant 220$ is occupied for all $j$. We let the positions of the agents evolve according to the ABM with $m=1$ for $q \in [0, 0.25, 0.5, 0.9]$.
Equation \eqref{eqn:ss_pde} is solved numerically with the same parameters under the same initial and boundary conditions as the ABM and the solution is averaged over the vertical extent of the domain (as in Equation \eqref{eqn:vert_ave}). The results are shown in Figure \ref{fig:ss_abm_pde_comparison}.

We note that the agents move less as we increase the adhesion strength. This agrees with our understanding of the ABM: the higher the adhesion the less likely agents are to break free of their neighbours and hence the less likely agents are to move. For $q=0$ the agreement between the ABM and PDE is excellent, as expected, since in the absence of adhesion, the movement process becomes a simple volume exclusion process \citep{liggett1999sis,simpson2007sic,simpson2009dpg,simpson2009mse,simpson2010cip,simpson2010mbc,cai2007mmw}. As we increase the adhesion strength to $q=0.25$ and $q=0.5$ slight deviations can be seen between the ABM and the PDE for large times. Nevertheless in general, the agreement remains promising. This disparity in agreement between the ABM and the PDE for higher adhesion strengths is due to the spatial correlations between lattice site occupancies in the ABM which are unaccounted for by the PDE approximation of the ABM.

\begin{figure}[t!]
\begin{center}
\includegraphics[width=0.60\textwidth]{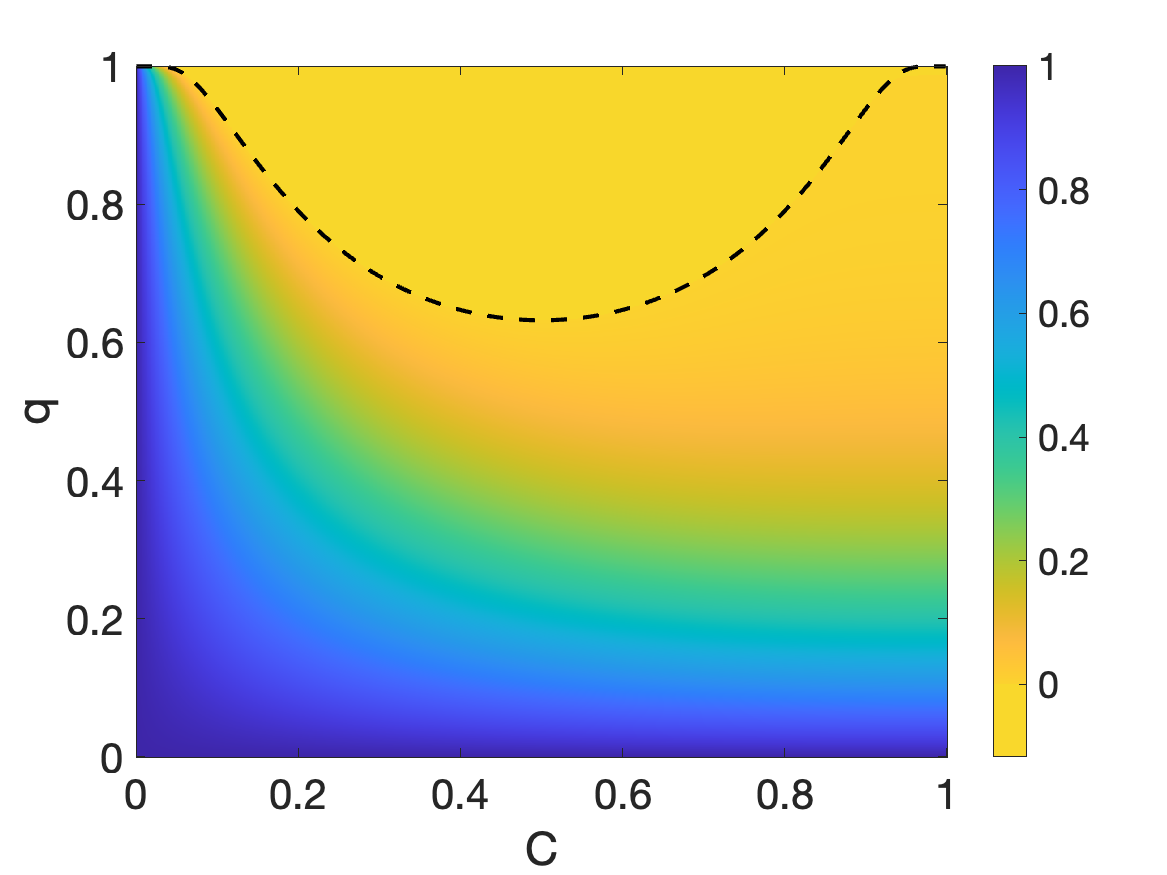}
\end{center}
\caption{Colourmap showing the dependence of the PDE diffusion coefficient $D(C)$ (see Equation \eqref{eqn:diff_coeff_ss}), on the adhesion strength $q$ and the species density $C$. The dashed curve is $q^\star$ (see Equation \eqref{eqn:bounded_region_D}) which bounds the region in $(C,q)$ space for which $D(C)<0$. The smallest value of $q$ for which $D(C)$ can be 0 is $q=1-\mathrm{e}^{-1}$ which permits $D(C)=0$ for $C=0.5$.}
\label{fig:heatmap}
\end{figure}

Increasing the adhesion strength further to $q=0.9$ the PDE becomes unstable as the macroscopic diffusion coefficient $D(C)$ defined in Equation \eqref{eqn:diff_coeff_ss} becomes negative for some regions of our space. The ABM shows minimal dispersion of agents which makes the density profiles at $t=0$, 100 and 1000 appear nearly indistinguishable from each other. The region in the $(C,q)$ space for which $D(C)<0$ is illustrated as a heatmap in Figure \ref{fig:heatmap}. The black dashed curve is defined by,

\begin{equation}\label{eqn:bounded_region_D}
q^\star=1-\exp\left\{\frac{1}{4C(C-1)}\right\},
\end{equation}

and bounds the region in $(C,q)$ space for which $D(C)<0$. A positive diffusion coefficient implies classical diffusion (in which mass descends gradients of mass) and a negative value implies aggregation (as mass climbs gradients of mass). The lowest value of adhesion strength which admits a non-positive diffusion coefficient is $q=1-\mathrm{e}^{-1}$. As a result, the PDE becomes unstable which suggests that adhesion-aided cell sorting into aggregation patterns such as engulfment cannot be observed using the PDEs.

In the next section, we develop a one-dimensional adhesion model for two species with cell-cell swapping. We derive the continuum model and show that it does a poor job of approximating the mean behaviour of the ABM when the adhesion between agents is strong. We then propose an alternative way to model the mean behaviour of the two species ABM which not only better represents the mean behaviour of the ABM than the continuum model but also allows us to replicate patterns observed in developmental biology.

\section{Two-species model}\label{sec:1d_abm}

In this section, we extend the single-species adhesion model to a population of two species, A and B. We model cell movement on a one-dimensional lattice, derive the corresponding one-dimensional PDEs and compare the average behaviour of the ABM to the numerical solution of the PDEs. We also explore the possibility of self-organisation into patterns as a result of cell sorting.

We initialise the lattice with $L$ sites of width $\Delta$. Let $m_A$ and $m_B$ be the rates of movement of species A and B, respectively, such that $m_A \delta t$ and $m_B \delta t$ are the probabilities that a type-A or type-B agent moves to either of its two neighbouring sites during a small time window of duration $\delta t$. If the target site is vacant the agent successfully moves. If, however, the target site is already occupied by another agent we attempt to swap the positions of the two agents with probability $\rho$ \citep{rajanoureen2023slc}. A successful swap results in the two neighbouring agents exchanging positions with each other and a failed swap results in the event being aborted in which case the two agents do not move.

To model cell-cell adhesion we introduce the adhesion strengths $p$, $q$ and $r$  ($0\leqslant p,q,r \leqslant 1$) where $p$ models adhesion between two type-A agents, $q$ models adhesion between two type-B agents and $r$ models adhesion between a type-A agent and a type-B agent. Defining $Z_{i}=\{i-1,i+1 \}$ as the set of positions of sites in the neighbourhood of site $i$, the probability that a type-A agent at site $i$ breaks its existing connections given that it is attempting to move is given by,

\begin{equation}\label{eqn:p_break_agent_A}
p_\text{break}^\text{agent}=(1-p)^{\mathrm{\sum}_{z\in Z_i} A_z} (1-r)^{\sum_{z\in Z_i}B_z},
\end{equation}

where $A_z$ is a binary indicator such that $A_z=1$ if the neighbour at position $z \in Z_i$ is a type-A agent and 0 otherwise. Similarly, $B_z$ is a binary indicator such that $B_z=1$ if the neighbouring agent at position $z \in Z_i$ is a type-B agent and 0 otherwise. Similarly, if the agent at site $i$ is a type-B agent then,

\begin{equation}\label{eqn:p_break_agent_B}
p_\text{break}^\text{agent}=(1-q)^{\mathrm{\sum}_{z\in Z_i} B_z} (1-r)^{\sum_{z\in Z_i}A_z}.
\end{equation}

If the target site is already occupied by another agent and the agents are attempting to swap positions then the probability that the target agent breaks its existing adhesive bonds given the agent at the target site is of type A is given by,

\begin{equation}\label{eqn:p_break_target_A}
p_\text{break}^\text{target}=(1-p)^{\mathrm{\sum}_{y\in Y_z} A_y} (1-r)^{\sum_{y \in Y_z}B_z},
\end{equation}

where $Y_z$ is the set of sites in the neighbourhood of the target site. Likewise, if the agent at the target site is a type-B agent,

\begin{equation}\label{eqn:p_break_target_B}
p_\text{break}^\text{target}=(1-q)^{\mathrm{\sum}_{y \in Y_z} B_y} (1-r)^{\sum_{y \in Y_z}A_z}.
\end{equation}

Altogether the probability of movement, given an agent and a target site have been chosen, is $p_{move} = p^\text{agent}_\text{break}(1-c_{y}) + \rho p^\text{agent}_\text{break} p^\text{target}_\text{break}c_{y}$ where $c_y \in \{0,1\}$ is the occupancy of the target site with 0 meaning the target site is vacant and 1 meaning that the site is occupied. The probabilities $p^\text{agent}_\text{break}$ and $p^\text{target}_\text{break}$ depend on whether the focal agent and, in the case of cell-cell swapping the agent being swapped with, are type-A or type-B agents, as indicated above in Equations  \eqref{eqn:p_break_agent_A}-\eqref{eqn:p_break_target_B}. Note that in the case of two swapping agents, we are modelling the breaking and reforming of the adhesive bonds between the two swapping agents explicitly although it would also be a valid choice to ignore the bonds between the two swapping agents altogether as they would be reformed following a successful swap.

\begin{figure}[t!]
\begin{center}

\subfigure[]{
\includegraphics[width=0.33\textwidth]{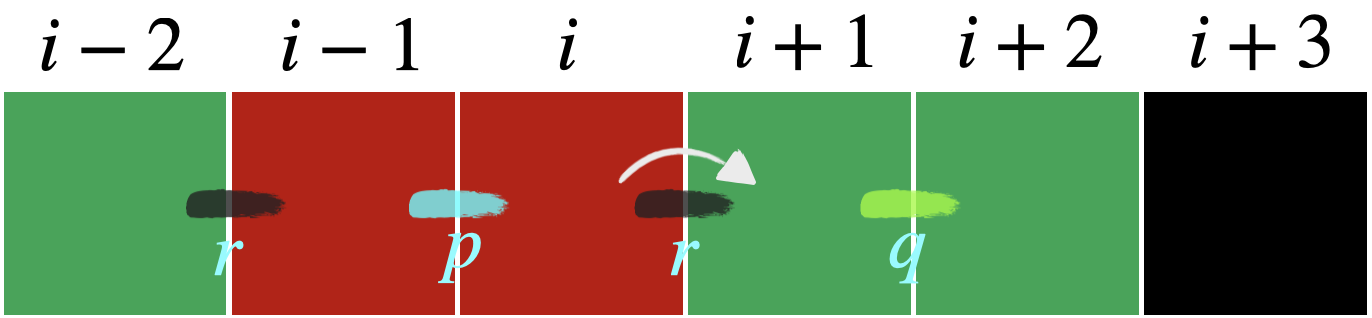}
\label{fig:1d_lattice_1}
}
\subfigure[]{
\includegraphics[width=0.33\textwidth]{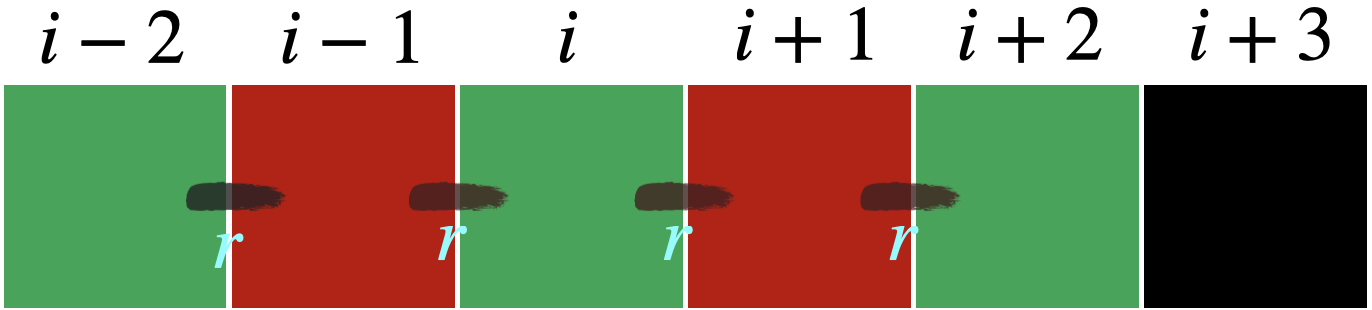}
\label{fig:1d_lattice_2}
}
\end{center}
\caption{Schematic illustrating an example of the two species ABM with adhesion. A red site represents a type-A agent, a green site represents a type-B agent and a black site represents an empty site. Adhesion strengths between neighbouring pairs of agents are shown as letters $p$, $q$ and $r$ alongside a dash representing a bond between a couple of agents. The initial configuration of the lattice is shown in \subref{fig:1d_lattice_1}. The arrow symbolises the red agent at site $i$ being chosen to attempt to move into the already occupied site $i+1$. In \subref{fig:1d_lattice_2} we show the result of a successful swap. Note that the labels denoting the adhesive bonds between pairs of agents have been updated in \subref{fig:1d_lattice_2}.}
\label{fig:1d_lattice_configs}
\end{figure}

In Figure \ref{fig:1d_lattice_configs}\subref{fig:1d_lattice_1} we show an example of an agent at site $i$ attempting to move into an already occupied site $i+1$. The focal agent is of type A with a type-A agent on its left (site $i-1$) and a type-B agent on its right (site $i+1$), thus $p^\text{break}_\text{agent} = (1-p)(1-r)$. The agent at the target site is a type-B agent with a type-A agent on its left (site $i$) and a type-B agent on its right (site $i+2$) and therefore $p^\text{break}_\text{target} = (1-q)(1-r)$. Hence, the probability of a successful swap between the pair of agents on this occasion is $p_\text{move} = \rho (1-p)(1-q)(1-r)^2$. If the swap is successful then the two agents make new connections with their neighbours following the move (see \subref{fig:1d_lattice_2}).

The agent-based model helps us understand and model interactions between individual agents. The average behaviour of a large number of cells can be found by averaging the ABM over a large number of repeats which can be computationally expensive. PDE-based continuum models provide us with a cheap and efficient way to approximate the mean behaviour of the discrete ABM. 
In the next section, we derive the continuum model corresponding to the one-dimensional ABM and we compare the two modelling paradigms to assess how well the continuum model approximates the mean behaviour of the ABM.

\subsection{One-dimensional continuum model}\label{sec:1D_model}

Let $\hat A_{i}(t)$ be the continuous approximation of the mean density of species A at the site with index $i$ at time $t$ and let $\hat B_{i}(t)$ be the same for species B. Then the continuous approximation of the mean densities of species A and species B, respectively, at the site $i$ at time $t+\delta t$, where $\delta t$ is infinitesimally small change in time, are given by,

\begin{align}\label{eqn:ts_mean_eq_1D_A}
& \hat A_{i} (t+\delta t) = \hat A_{i} + \frac{m_A}{2} \delta t \hat A_{i-1}(1-\hat A_{i}-\hat B_{i})(1-p)^{\Sigma_{z \in Z_{i-1}} \hat A_z} (1-r)^{\Sigma_{z \in Z_{i-1}} \hat B_z} \nonumber \\
& \quad +\frac{m_A}{2} \delta t \hat A_{i+1}(1-\hat A_{i}-\hat B_{i})(1-p)^{\Sigma_{z \in Z_{i+1}} \hat A_z} (1-r)^{\Sigma_{z \in Z_{i+1}} \hat B_z} \nonumber \\ 
& \quad - \Big(\frac{m_A}{2}\delta t(1-\hat A_{i-1}-\hat B_{i-1})(1-p)^{\Sigma_{z \in Z_{i}} \hat A_z}(1-r)^{\Sigma_{z \in Z_{i}} \hat B_z} \nonumber \\
& \quad + \frac{m_A}{2}\delta t(1-\hat A_{i+1}-\hat B_{i+1})(1-p)^{\Sigma_{z \in Z_{i}} \hat A_z}(1-r)^{\Sigma_{z \in Z_{i}} \hat B_z} \Big) \nonumber \\ 
& \quad + \frac{(m_A+m_B)}{2}\delta t \rho \hat A_{i-1}\hat B_{i}(1-p)^{\Sigma_{z \in Z_{i-1}} \hat A_z}(1-q)^{\Sigma_{y \in Y_{i}} \hat B_y}(1-r)^{\Sigma_{z \in Z_{i-1}} \hat B_z}(1-r)^{\Sigma_{y \in Y_{i}} \hat A_y}\nonumber \\
& \quad + \frac{(m_A+m_B)}{2}\delta t \rho \hat A_{i+1}\hat B_{i}(1-p)^{\Sigma_{z \in Z_{i+1}} \hat A_z}(1-q)^{\Sigma_{y \in Y_{i}} \hat B_y}(1-r)^{\Sigma_{z \in Z_{i+1}} \hat B_z}(1-r)^{\Sigma_{y \in Y_{i}} \hat A_y} \nonumber \\
& \quad - \frac{(m_A+m_B)}{2}\delta t \rho \hat A_{i}\hat B_{i-1}(1-p)^{\Sigma_{z \in Z_{i}} \hat A_z}(1-q)^{\Sigma_{y \in Y_{i-1}} \hat B_y}(1-r)^{\Sigma_{z \in Z_{i}} \hat B_z}(1-r)^{\Sigma_{y \in Y_{i-1}} \hat A_y}\nonumber \\
& \quad - \frac{(m_A+m_B)}{2}\delta t \rho \hat A_{i}\hat B_{i+1}(1-p)^{\Sigma_{z \in Z_{i}} \hat A_z}(1-q)^{\Sigma_{y \in Y_{i+1}} \hat B_y}(1-r)^{\Sigma_{z \in Z_{i}} \hat B_z}(1-r)^{\Sigma_{y \in Y_{i+1}}\hat A_y},
\end{align}

\begin{align}\label{eqn:ts_mean_eq_1D_B}
& \hat B_{i} (t+\delta t) = \hat B_{i} + \frac{m_B}{2} \delta t \hat B_{i-1}(1-\hat B_{i}-\hat A_{i})(1-q)^{\Sigma_{z \in Z_{i-1}} \hat B_z} (1-r)^{\Sigma_{z \in Z_{i-1}} \hat A_z} \nonumber \\
& \quad +\frac{m_B}{2} \delta t \hat B_{i+1}(1-\hat B_{i}-\hat A_{i})(1-q)^{\Sigma_{z \in Z_{i+1}} \hat B_z} (1-r)^{\Sigma_{z \in Z_{i+1}} \hat A_z} \nonumber \\ 
& \quad - \Big(\frac{m_B}{2}\delta t(1-\hat B_{i-1}-\hat A_{i-1})(1-q)^{\Sigma_{z \in Z_{i}} \hat B_z}(1-r)^{\Sigma_{z \in Z_{i}} \hat A_z} \nonumber \\
& \quad + \frac{m_B}{2}\delta t(1-\hat B_{i+1}-\hat A_{i+1})(1-q)^{\Sigma_{z \in Z_{i}} \hat B_z}(1-r)^{\Sigma_{z \in Z_{i}} \hat A_z} \Big) \nonumber \\ 
& \quad + \frac{(m_B+m_A)}{2}\delta t \rho \hat B_{i-1}\hat A_{i}(1-q)^{\Sigma_{z \in Z_{i-1}} \hat B_z}(1-p)^{\Sigma_{y \in Y_{i}} \hat A_y}(1-r)^{\Sigma_{z \in Z_{i-1}} \hat A_z}(1-r)^{\Sigma_{y \in Y_{i}} \hat B_y}\nonumber \\
& \quad + \frac{(m_B+m_A)}{2}\delta t \rho \hat B_{i+1}\hat A_{i}(1-q)^{\Sigma_{z \in Z_{i+1}} \hat B_z}(1-p)^{\Sigma_{y \in Y_{i}} \hat A_y}(1-r)^{\Sigma_{z \in Z_{i+1}} \hat A_z}(1-r)^{\Sigma_{y \in Y_{i}} \hat B_y} \nonumber \\
& \quad - \frac{(m_B+m_A)}{2}\delta t \rho \hat B_{i}\hat A_{i-1}(1-q)^{\Sigma_{z \in Z_{i}} \hat B_z}(1-p)^{\Sigma_{y \in Y_{i-1}} \hat A_y}(1-r)^{\Sigma_{z \in Z_{i}} \hat A_z}(1-r)^{\Sigma_{y \in Y_{i-1}} \hat B_y}\nonumber \\
& \quad - \frac{(m_B+m_A)}{2}\delta t \rho \hat B_{i}\hat A_{i+1}(1-q)^{\Sigma_{z \in Z_{i}} \hat B_z}(1-p)^{\Sigma_{y \in Y_{i+1}} \hat A_y}(1-r)^{\Sigma_{z \in Z_{i}} \hat A_z}(1-r)^{\Sigma_{y \in Y_{i+1}}\hat B_y}.
\end{align}

Taylor expanding terms in Equations \eqref{eqn:ts_mean_eq_1D_A} and \eqref{eqn:ts_mean_eq_1D_B} around the site $i$ and taking $\Delta, \; \delta t \to 0$, such that $\Delta^2/\delta t$ remains constant, leads to the coupled one-dimensional PDEs,

\begin{align}\label{eqn:ts_pde_1D_A}
\D A t &= \frac{\partial }{\partial x} \Bigg[D_A\Big(D_1(A,B) \D B x +D_2(A,B)\D A x \Big) + (D_A+D_B)\rho \Big(D_3(A,B) \D A x- D_4(A,B) \D B x \Big)\Bigg],
\end{align}

\begin{align}\label{eqn:ts_pde_1D_B}
\D B t &= \frac{\partial }{\partial x} \Bigg[D_B \Big(D_5(A,B) \D A x + D_6(A,B)\D B x\Big) + (D_A+D_B)\rho \Big( D_7(A,B) \D B x - D_8(A,B) \D A x \Big) \Bigg],
\end{align}

where,

\begin{equation}
D_A= \lim_{\substack{\Delta \to 0 \\ \delta t \to 0}}\frac{m_A\Delta^2}{2\delta t}, \qquad D_B= \lim_{\substack{\Delta \to 0 \\ \delta t \to 0}}\frac{m_B\Delta^2}{2\delta t},
\end{equation}

and,

\begin{align}
D_1(A,B)&=(1-p)^{2A}(1-r)^{2B}(2 \ln(1-r)(1-A-B)+1)A,\\
D_2(A,B)&=(1-p)^{2A}(1-r)^{2B}(2\ln(1-p)A(1-A-B)+1-B),\\
D_3(A,B)&=(1-p)^{2A}(1-q)^{2B}(1-r)^{2A+2B}B(1-2\ln(1-r)A+2\ln(1-p)A),\\
D_4(A,B)&=(1-p)^{2A}(1-q)^{2B}(1-r)^{2A+2B}A(1-2\ln(1-q)B+2\ln(1-r)B),\\
D_5(A,B)&=(1-q)^{2B}(1-r)^{2A}(2 \ln(1-r)(1-A-B)+1)B,\\
D_6(A,B)&=(1-q)^{2B}(1-r)^{2A}(2\ln(1-q)B(1-A-B)+1-A),\\
D_7(A,B)&=(1-q)^{2B}(1-p)^{2A}(1-r)^{2A+2B}A(1-2\ln(1-r)B+2\ln(1-q)B),\\
D_8(A,B)&=(1-q)^{2B}(1-p)^{2A}(1-r)^{2A+2B}B(1-2\ln(1-p)A+2\ln(1-r)A).
\end{align}

\begin{figure}[t!]
\begin{center}

\subfigure[]{
\includegraphics[width=0.31\textwidth]{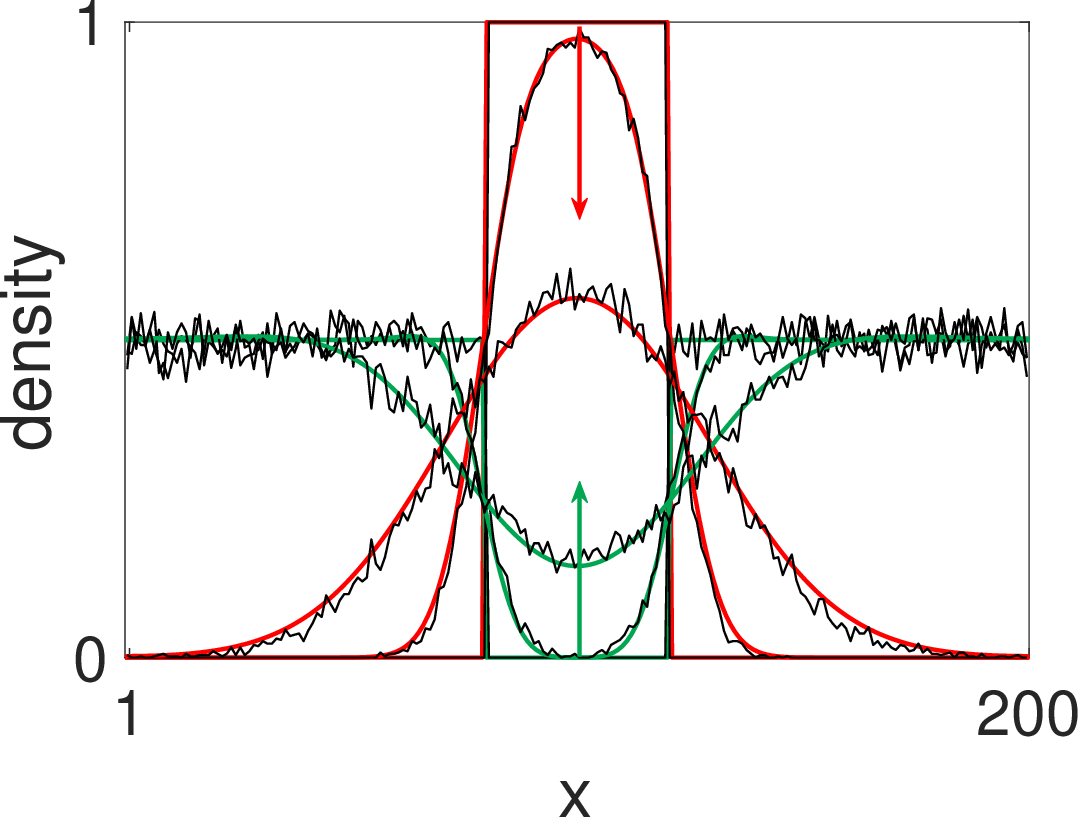}
\label{fig:ts_1D_p=0_q=0.25_r=0_rho=0.25}
}
\subfigure[]{
\includegraphics[width=0.31\textwidth]{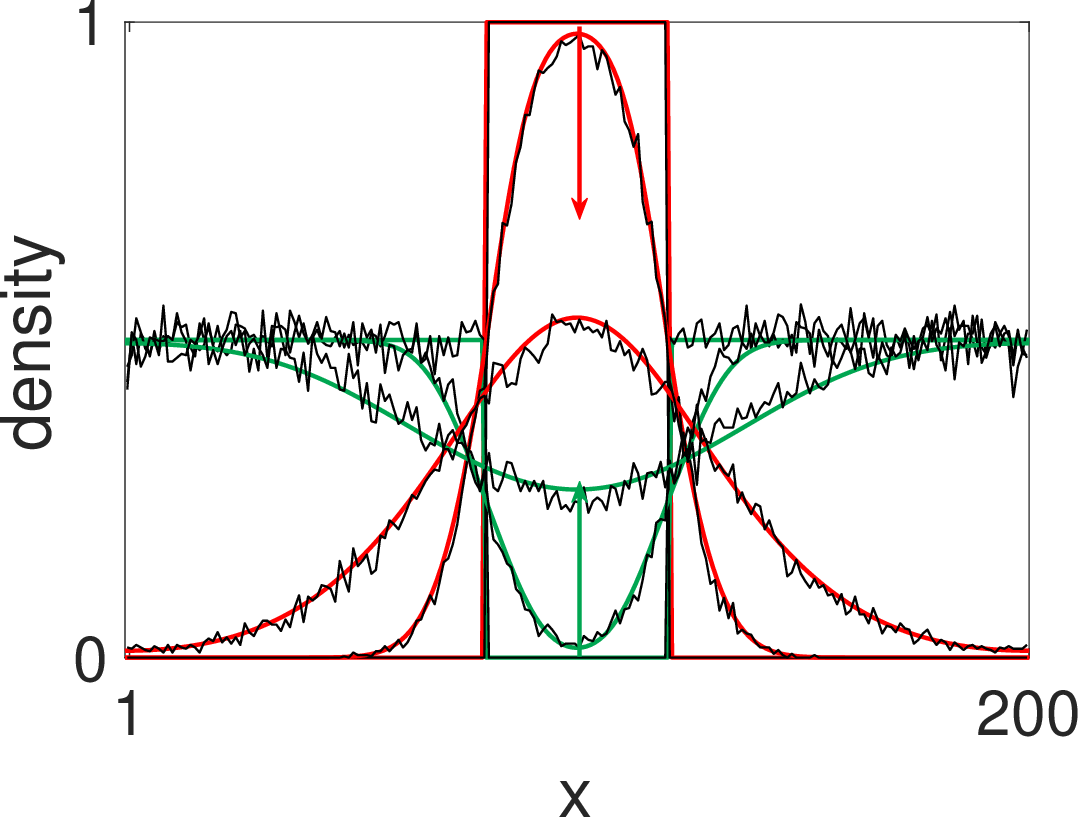}
\label{fig:ts_1D_p=0.25_q=0.25_r=0_rho=0.75}
}
\subfigure[]{
\includegraphics[width=0.31\textwidth]{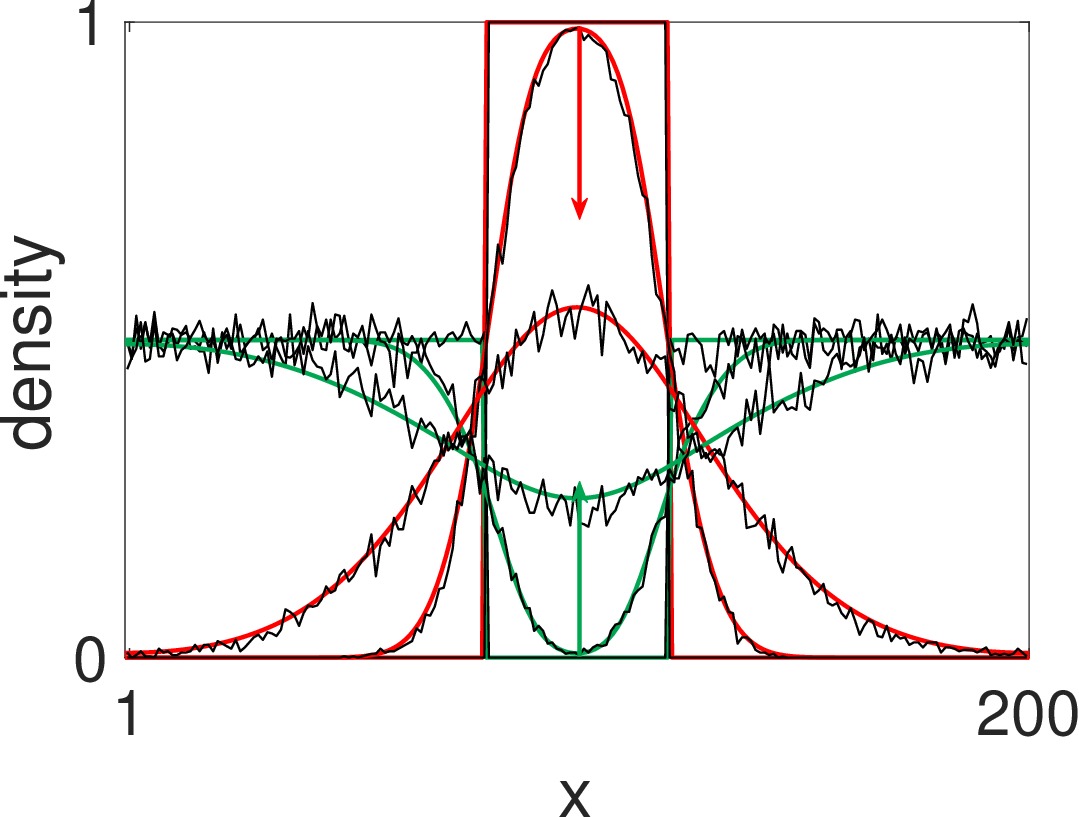}
\label{fig:ts_1D_p=0.25_q=0.25_r=0.25_rho=1}
}
\end{center}
\caption{Comparison between the average two-species ABM and the PDEs \eqref{eqn:ts_pde_1D_A} and \eqref{eqn:ts_pde_1D_B} for set of parameters \subref{fig:ts_1D_p=0_q=0.25_r=0_rho=0.25} $p=r=0, \; q=0.25,\; \rho=0.25$ \subref{fig:ts_1D_p=0.25_q=0.25_r=0_rho=0.75}, $p=q=0.25, \; r=0, \; \rho=0.75$ and \subref{fig:ts_1D_p=0.25_q=0.25_r=0.25_rho=1} $p=q=r=0.25, \; \rho=1$. For all the three figures presented here the movement rates are $m_A=m_B=1$ and we impose zero-flux boundary conditions in both the ABM and the PDE. The ABM is shown as black noisy curves and the PDE solution is presented in red for species A and green for species B. The domain of length $L=200$ is initialised such that the central 20\% of it has density 1 of type A agents and the rest of the domain is initialised with type-B agents with a density of 0.5. The ABM is averaged over 500 repeats and we show the density of species A and species B at $t=$ 0, 100 and 1,000. The red and green arrows show the direction of increasing time.}
\label{fig:ts_1D_abm_pde_comparison_good}
\end{figure}

In order to compare our ABM and PDE solutions, we consider a domain of length $L=200$ and initialise it such that the central 20\% of the domain has density 1 of type A agents and the rest of the has density 0.5 of type-B agents, i.e.,

\begin{equation}
c_A=\begin{cases}
1 & \text{if } 81 \leqslant x \leqslant 120,\\
0 & \text{otherwise},
\end{cases}
\end{equation}

and,

\begin{equation}
c_B=\begin{cases}
0.5 & \text{if } 1 \leqslant x \leqslant 80 \text{ and } 121 \leqslant x \leqslant 200,\\
0 & \text{otherwise}.
\end{cases}
\end{equation}

\begin{figure}[t!]
\begin{center}
\subfigure[]{
\includegraphics[width=0.45\textwidth]{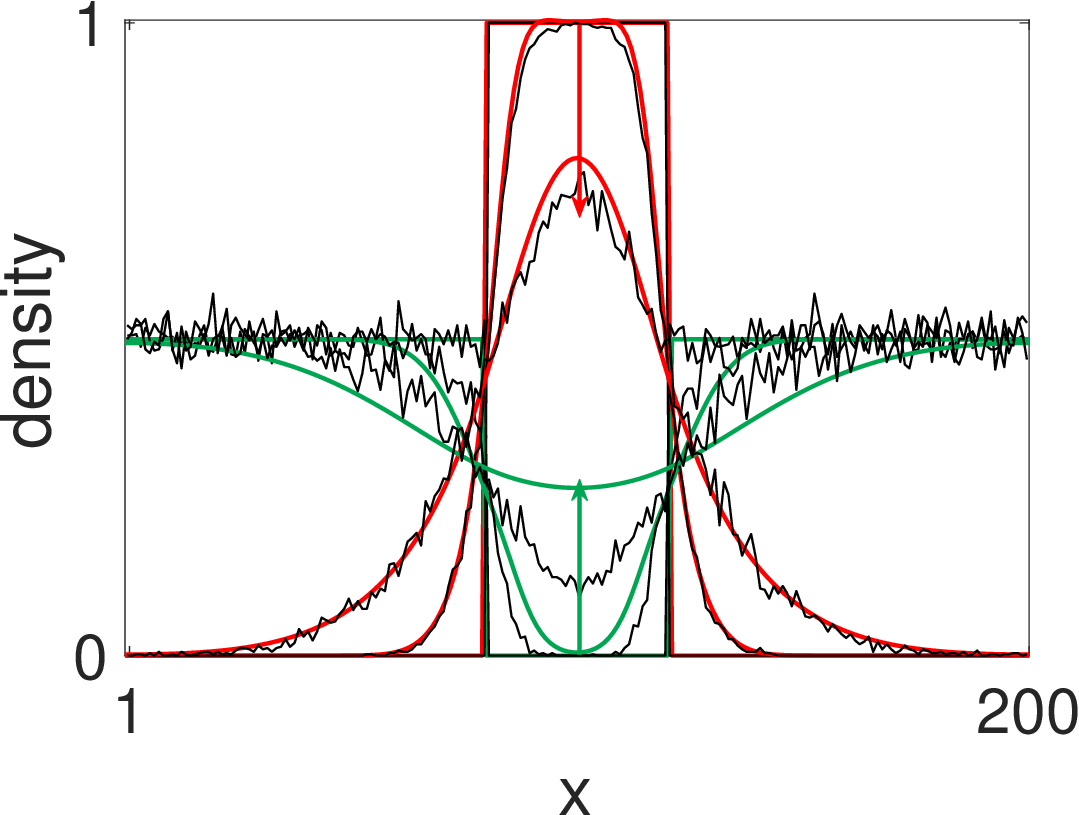}
\label{fig:one_dimensional_ts_p_0.5_q_0_r_0.25_rho_0.5}
}
\subfigure[]{
\includegraphics[width=0.45\textwidth]{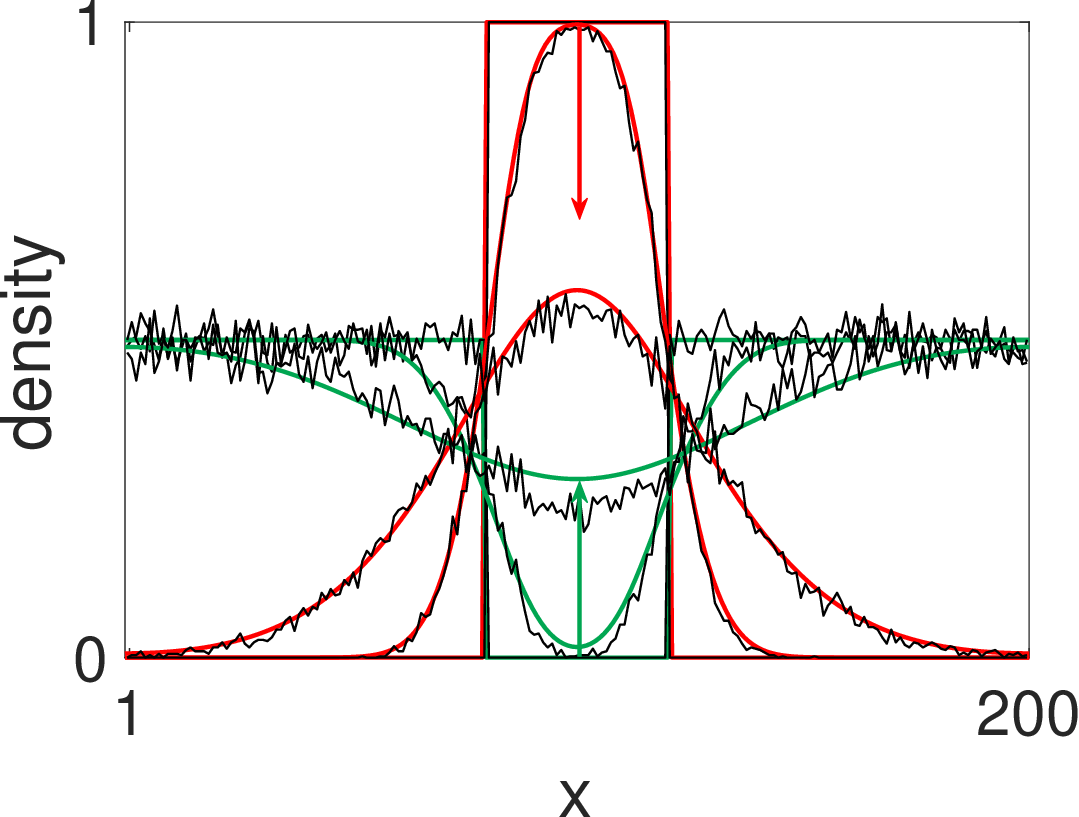}
\label{fig:one_dimensional_ts_p_0.25_q_0_r_0.25_rho_0.75}
}

\subfigure[]{
\includegraphics[width=0.45\textwidth]{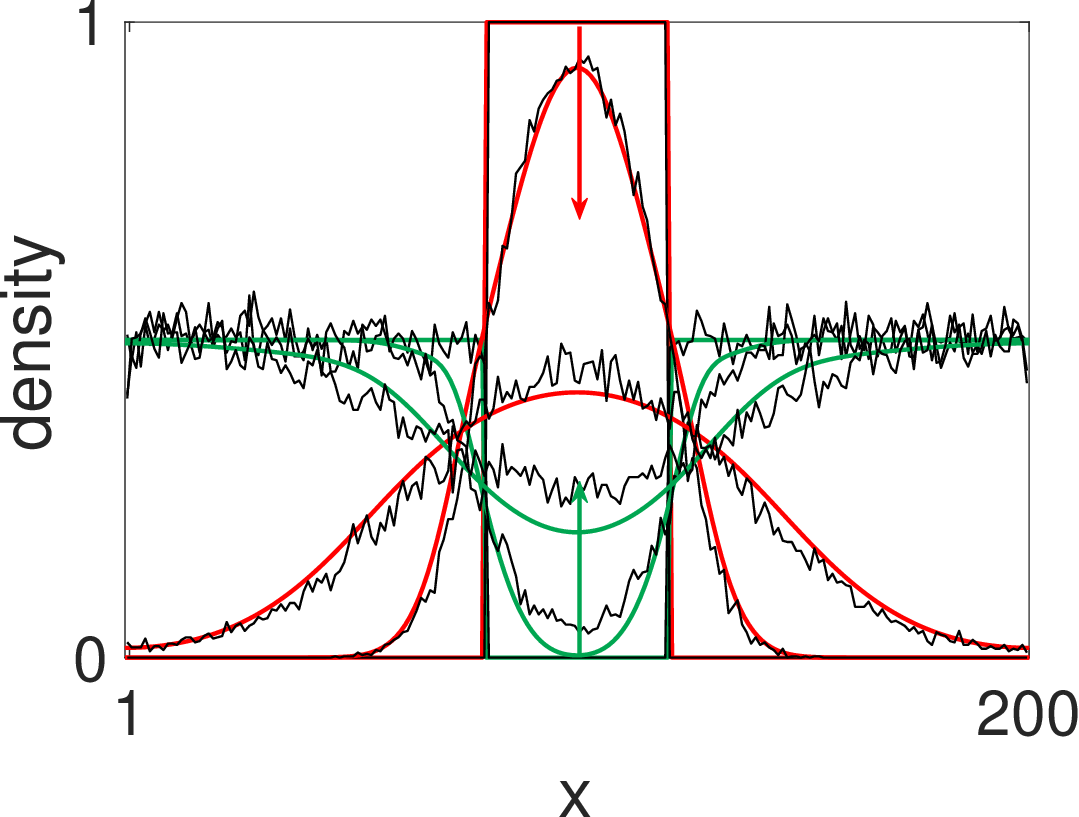}
\label{fig:one_dimensional_ts_p_0_q_0.5_r_0_rho_1}
}
\subfigure[]{
\includegraphics[width=0.45\textwidth]{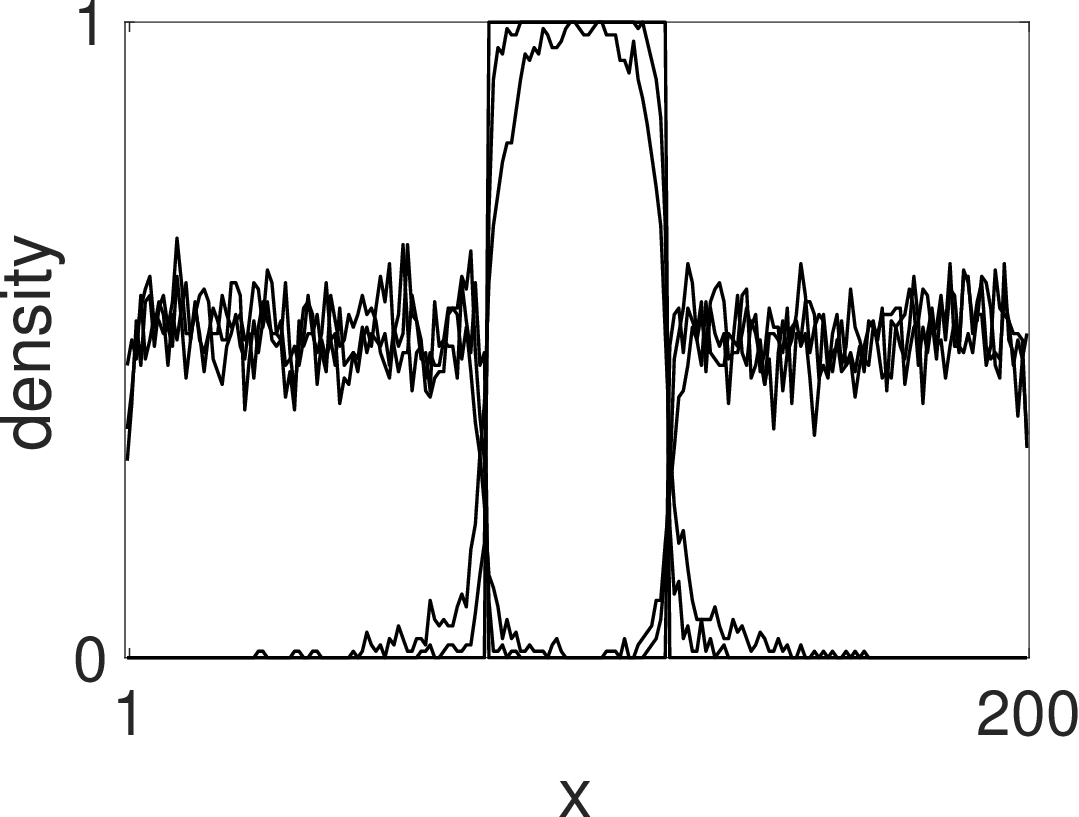}
\label{fig:one_dimensional_ts_p_0.9_q_0.8_r_0_rho_1}
}
\end{center}
\caption{Poor agreement between the two-species ABM and the PDEs \eqref{eqn:ts_pde_1D_A} and \eqref{eqn:ts_pde_1D_B}. Figure descriptions are as in Figure \ref{fig:ts_1D_abm_pde_comparison_good}. Parameter values used here are: \subref{fig:one_dimensional_ts_p_0.5_q_0_r_0.25_rho_0.5} $p=0.50, \; q=0, \; r=0.25, \; \rho=0.50$ , \subref{fig:one_dimensional_ts_p_0.25_q_0_r_0.25_rho_0.75} $p=r=0.25, \; q=0, \; \rho=0.75$,  \subref{fig:one_dimensional_ts_p_0_q_0.5_r_0_rho_1} $p=0, \; q=0.5, \; r=0, \; \rho=1$ and \subref{fig:one_dimensional_ts_p_0.9_q_0.8_r_0_rho_1} $p=0.90, \; q=0.80, \; r=0, \; \rho=1$. The red and green arrows show the direction of increasing time in panels \subref{fig:one_dimensional_ts_p_0.5_q_0_r_0.25_rho_0.5}-\subref{fig:one_dimensional_ts_p_0_q_0.5_r_0_rho_1}.}
\label{fig:ts_1D_abm_pde_comparison_bad}
\end{figure}

We let the positions of the agents evolve according to the two-species ABM and solve the corresponding PDEs numerically with zero flux boundary conditions. In Figure \ref{fig:ts_1D_abm_pde_comparison_good} we present the average density of agents in ABM and the continuum model at $t=$ 0, 100 and 1,000 for three different sets of low adhesion strengths and a range of swapping probabilities. Good agreement between the average ABM and the continuum model can be observed in each case. In Figure \ref{fig:ts_1D_abm_pde_comparison_bad}\subref{fig:one_dimensional_ts_p_0.5_q_0_r_0.25_rho_0.5}-\subref{fig:one_dimensional_ts_p_0_q_0.5_r_0_rho_1} we see that the agreement is not as good for some other adhesion strengths due to correlations between sites' occupancies which the continuum model ignores. Furthermore, for very high self-adhesion (panel \subref{fig:one_dimensional_ts_p_0.9_q_0.8_r_0_rho_1}) in the ABM, the overall movement of agents is significantly reduced as a result of fewer bonds breaking between pairs of neighbouring agents. The PDE, on the other hand, is unstable for such strong adhesion and consequently we have not been able to plot its solution. In this particular case the diffusion/cross-diffusion coefficients $D_1(A,B), \; D_3(A,B), \; D_4(A,B), \; D_6(A,B)$ and $D_7(A,b)$ all become negative which implies the potential for aggregation. While the PDE mimics the average behaviour of the ABM well for some parameter regimes, it fails to represent the true average dynamics of the ABM in other cases. In addition, the continuum model is not an appropriate approach for modelling patterns that emerge as a result of adhesion-mediated cell sorting.

Instead, we can use stochastic mean equations  (SMEs) to represent the mean behaviour of the agents \citep{thompson2012mcm, anguige2008omc}. We can derive the SMEs directly from Equation \eqref{eqn:ts_mean_eq_1D_A} and \eqref{eqn:ts_mean_eq_1D_B} with some algebraic manipulation and taking $\delta t \to 0$ which gives us an ODE for each lattice site $i$ making a system of $L$ equations (one for each site) for each species:

\begin{align}\label{eqn:sme_A}
\frac{d \hat A_{i} (t)}{d t} &= \frac{m_A}{2} \delta t \hat A_{i-1}(1-\hat A_{i}-\hat B_{i})(1-p)^{\Sigma_{z \in Z_{i-1}} \hat A_z} (1-r)^{\Sigma_{z \in Z_{i-1}} \hat B_z} \nonumber \\
& \quad +\frac{m_A}{2} \delta t \hat A_{i+1}(1-\hat A_{i}-\hat B_{i})(1-p)^{\Sigma_{z \in Z_{i+1}} \hat A_z} (1-r)^{\Sigma_{z \in Z_{i+1}} \hat B_z} \nonumber \\ 
& \quad - \Big(\frac{m_A}{2}\delta t(1-\hat A_{i-1}-\hat B_{i-1})(1-p)^{\Sigma_{z \in Z_{i}} \hat A_z}(1-r)^{\Sigma_{z \in Z_{i}} \hat B_z} \nonumber \\
& \quad + \frac{m_A}{2}\delta t(1-\hat A_{i+1}-\hat B_{i+1})(1-p)^{\Sigma_{z \in Z_{i}} \hat A_z}(1-r)^{\Sigma_{z \in Z_{i}} \hat B_z} \Big) \nonumber \\ 
& \quad + \frac{(m_A+m_B)}{2}\delta t \rho \hat A_{i-1}\hat B_{i}(1-p)^{\Sigma_{z \in Z_{i-1}} \hat A_z}(1-q)^{\Sigma_{y \in Y_{i}} \hat B_y}(1-r)^{\Sigma_{z \in Z_{i-1}} \hat B_z}(1-r)^{\Sigma_{y \in Y_{i}} \hat A_y}\nonumber \\
& \quad + \frac{(m_A+m_B)}{2}\delta t \rho \hat A_{i+1}\hat B_{i}(1-p)^{\Sigma_{z \in Z_{i+1}} \hat A_z}(1-q)^{\Sigma_{y \in Y_{i}} \hat B_y}(1-r)^{\Sigma_{z \in Z_{i+1}} \hat B_z}(1-r)^{\Sigma_{y \in Y_{i}} \hat A_y} \nonumber \\
& \quad - \frac{(m_A+m_B)}{2}\delta t \rho \hat A_{i}\hat B_{i-1}(1-p)^{\Sigma_{z \in Z_{i}} \hat A_z}(1-q)^{\Sigma_{y \in Y_{i-1}} \hat B_y}(1-r)^{\Sigma_{z \in Z_{i}} \hat B_z}(1-r)^{\Sigma_{y \in Y_{i-1}} \hat A_y}\nonumber \\
& \quad - \frac{(m_A+m_B)}{2}\delta t \rho \hat A_{i}\hat B_{i+1}(1-p)^{\Sigma_{z \in Z_{i}} \hat A_z}(1-q)^{\Sigma_{y \in Y_{i+1}} \hat B_y}(1-r)^{\Sigma_{z \in Z_{i}} \hat B_z}(1-r)^{\Sigma_{y \in Y_{i+1}}\hat A_y},
\end{align}

\begin{align}\label{eqn:sme_B}
\frac{d \hat B_{i} (t)}{d t} &= \frac{m_B}{2} \delta t \hat B_{i-1}(1-\hat B_{i}-\hat A_{i})(1-q)^{\Sigma_{z \in Z_{i-1}} \hat B_z} (1-r)^{\Sigma_{z \in Z_{i-1}} \hat A_z} \nonumber \\
& \quad +\frac{m_B}{2} \delta t \hat B_{i+1}(1-\hat B_{i}-\hat A_{i})(1-q)^{\Sigma_{z \in Z_{i+1}} \hat B_z} (1-r)^{\Sigma_{z \in Z_{i+1}} \hat A_z} \nonumber \\ 
& \quad - \Big(\frac{m_B}{2}\delta t(1-\hat B_{i-1}-\hat A_{i-1})(1-q)^{\Sigma_{z \in Z_{i}} \hat B_z}(1-r)^{\Sigma_{z \in Z_{i}} \hat A_z} \nonumber \\
& \quad + \frac{m_B}{2}\delta t(1-\hat B_{i+1}-\hat A_{i+1})(1-q)^{\Sigma_{z \in Z_{i}} \hat B_z}(1-r)^{\Sigma_{z \in Z_{i}} \hat A_z} \Big) \nonumber \\ 
& \quad + \frac{(m_B+m_A)}{2}\delta t \rho \hat B_{i-1}\hat A_{i}(1-q)^{\Sigma_{z \in Z_{i-1}} \hat B_z}(1-p)^{\Sigma_{y \in Y_{i}} \hat A_y}(1-r)^{\Sigma_{z \in Z_{i-1}} \hat A_z}(1-r)^{\Sigma_{y \in Y_{i}} \hat B_y}\nonumber \\
& \quad + \frac{(m_B+m_A)}{2}\delta t \rho \hat B_{i+1}\hat A_{i}(1-q)^{\Sigma_{z \in Z_{i+1}} \hat B_z}(1-p)^{\Sigma_{y \in Y_{i}} \hat A_y}(1-r)^{\Sigma_{z \in Z_{i+1}} \hat A_z}(1-r)^{\Sigma_{y \in Y_{i}} \hat B_y} \nonumber \\
& \quad - \frac{(m_B+m_A)}{2}\delta t \rho \hat B_{i}\hat A_{i-1}(1-q)^{\Sigma_{z \in Z_{i}} \hat B_z}(1-p)^{\Sigma_{y \in Y_{i-1}} \hat A_y}(1-r)^{\Sigma_{z \in Z_{i}} \hat A_z}(1-r)^{\Sigma_{y \in Y_{i-1}} \hat B_y}\nonumber \\
& \quad - \frac{(m_B+m_A)}{2}\delta t \rho \hat B_{i}\hat A_{i+1}(1-q)^{\Sigma_{z \in Z_{i}} \hat B_z}(1-p)^{\Sigma_{y \in Y_{i+1}} \hat A_y}(1-r)^{\Sigma_{z \in Z_{i}} \hat A_z}(1-r)^{\Sigma_{y \in Y_{i+1}}\hat B_y}.
\end{align}

\begin{figure}[t!]
\begin{center}
\subfigure[]{
\includegraphics[width=0.31\textwidth]{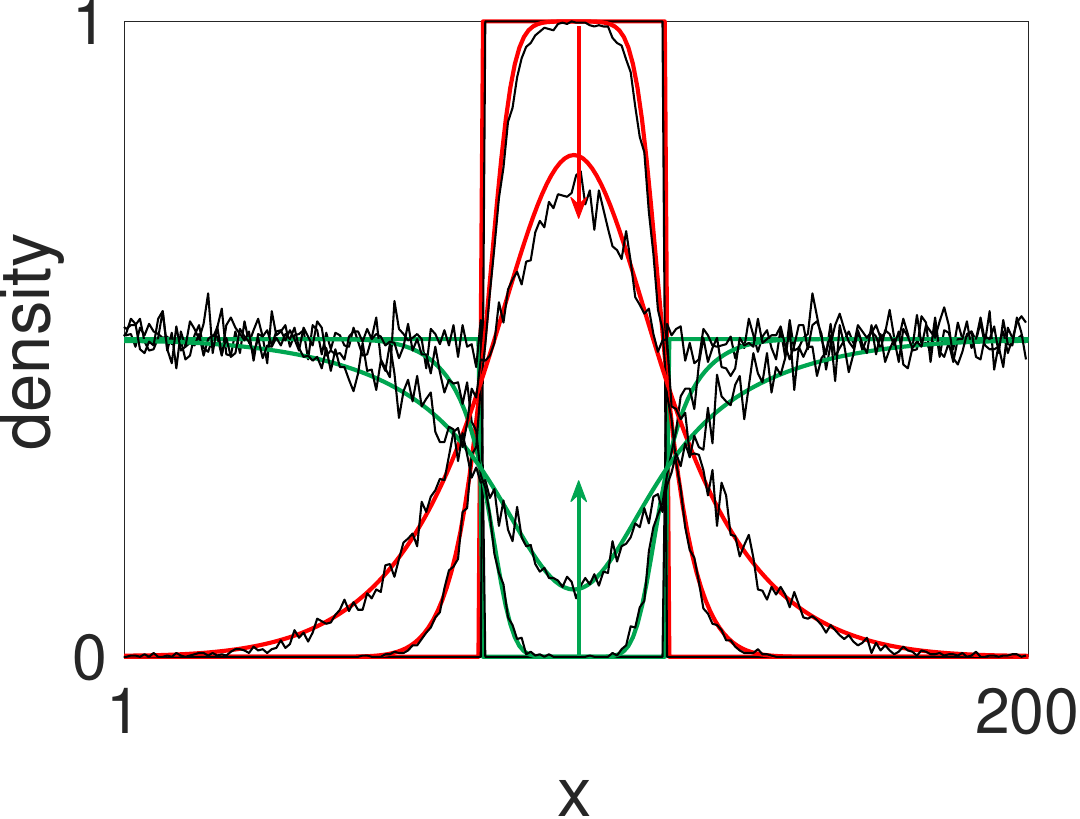}
\label{fig:one_dimensional_ts_sme_p_0.5_q_0_r_0.25_rho_0.5}
}
\subfigure[]{
\includegraphics[width=0.31\textwidth]{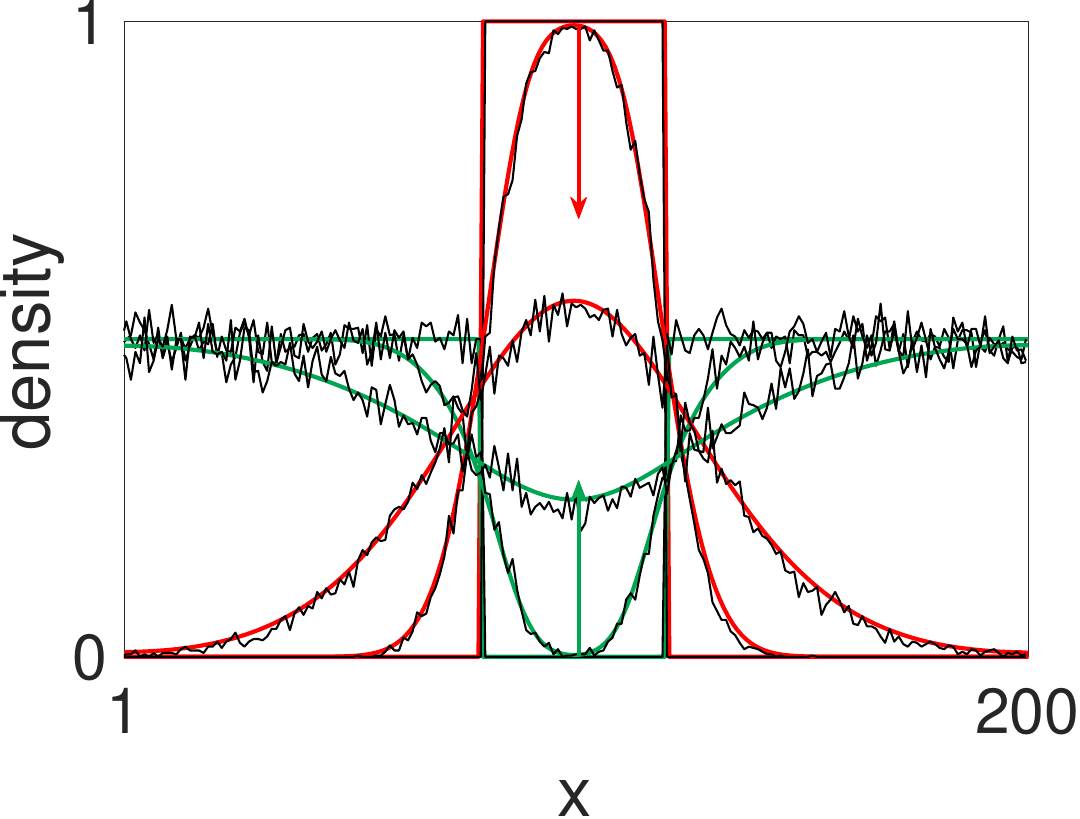}
\label{fig:one_dimensional_ts_sme_p_0.25_q_0_r_0.25_rho_0.75}
}
\subfigure[]{
\includegraphics[width=0.31\textwidth]{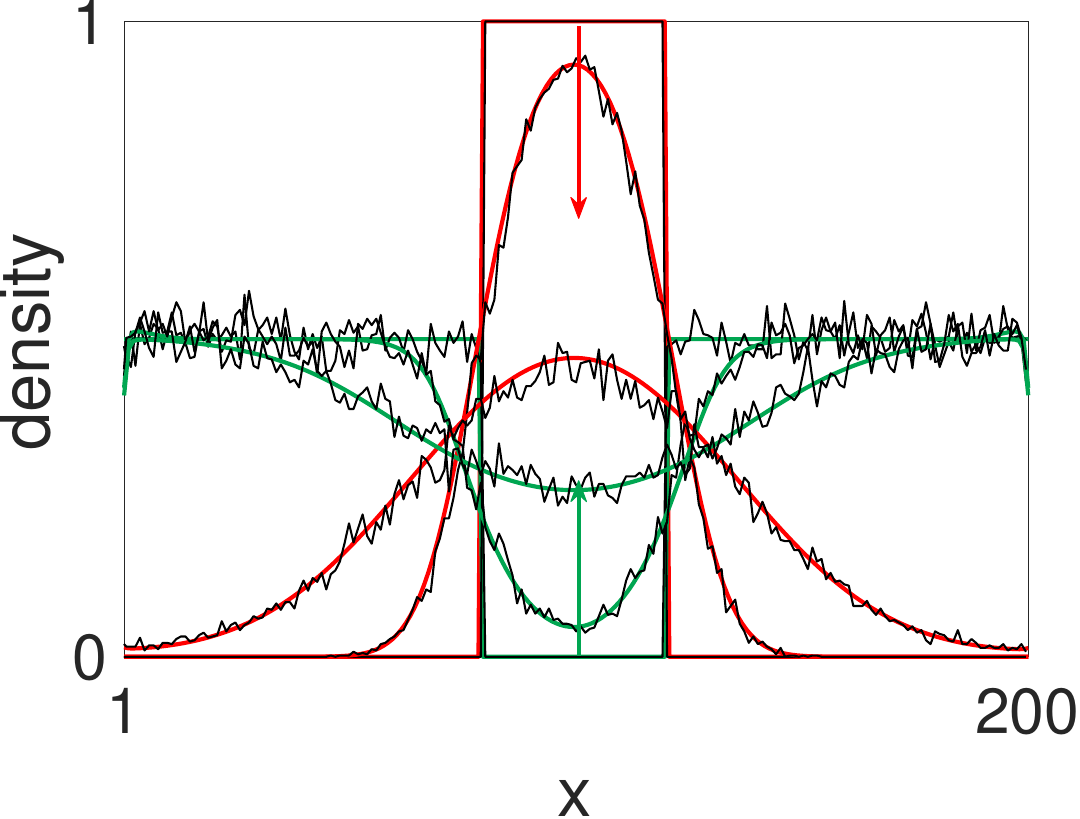}
\label{fig:one_dimensional_ts_sme_p_0_q_0.5_r_0_rho_1}
}
\end{center}
\caption{Comparison between the average two-species ABM and the numerical solution of the SMEs \eqref{eqn:sme_A} and \eqref{eqn:sme_B} for the same parameters as Figure \ref{fig:ts_1D_abm_pde_comparison_bad}\subref{fig:one_dimensional_ts_p_0.5_q_0_r_0.25_rho_0.5}-\subref{fig:one_dimensional_ts_p_0_q_0.5_r_0_rho_1} shown here in panels \subref{fig:one_dimensional_ts_sme_p_0.5_q_0_r_0.25_rho_0.5}-\subref{fig:one_dimensional_ts_sme_p_0_q_0.5_r_0_rho_1}, respectively. The initial condition is the same as Figure \ref{fig:ts_1D_abm_pde_comparison_bad}.}
\label{fig:ts_1D_abm_sme_comparison}
\end{figure}

\begin{figure}[t!]
\begin{center}
\subfigure[]{
\includegraphics[width=0.31\textwidth]{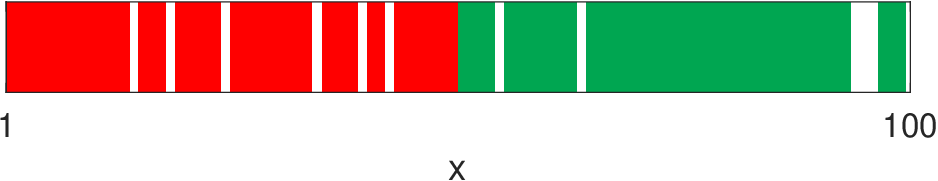}
\label{fig:pattern_abm_p_0.6_q_0.55_r_0.5_rho_1_t_0}
}
\subfigure[]{
\includegraphics[width=0.31\textwidth]{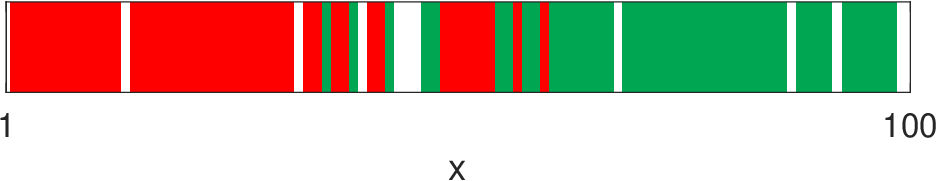}
\label{fig:pattern_abm_p_0.6_q_0.55_r_0.5_rho_1_t_1000}
}
\subfigure[]{
\includegraphics[width=0.31\textwidth]{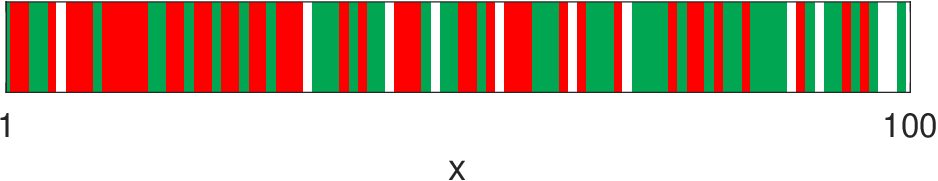}
\label{fig:pattern_abm_p_0.6_q_0.55_r_0.5_rho_1_t_100000}
}

\subfigure[]{
\includegraphics[width=0.31\textwidth]{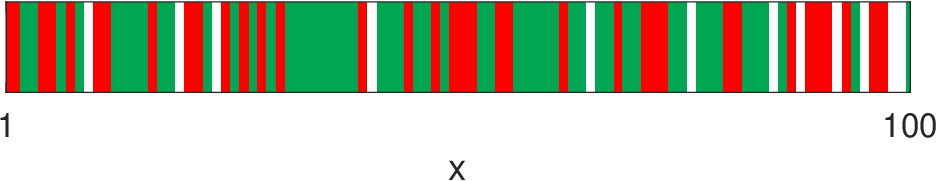}
\label{fig:pattern_abm_p_0.9_q_0.3_r_0_rho_1_t_0}
}
\subfigure[]{
\includegraphics[width=0.31\textwidth]{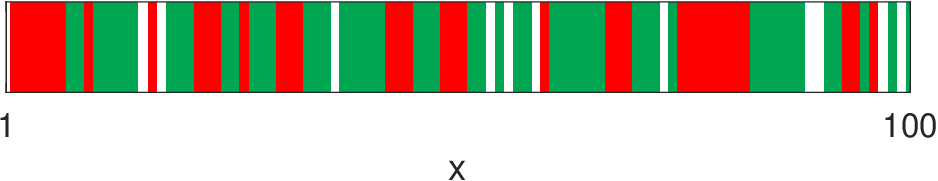}
\label{fig:pattern_abm_p_0.9_q_0.3_r_0_rho_1_t_1000}
}
\subfigure[]{
\includegraphics[width=0.31\textwidth]{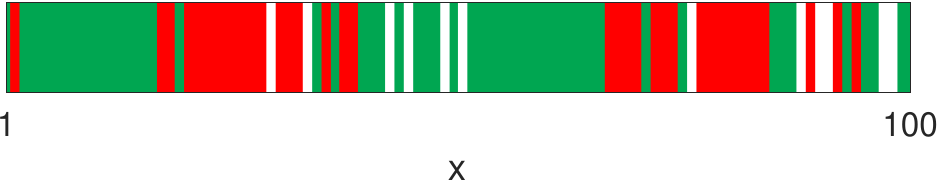}
\label{fig:pattern_abm_p_0.9_q_0.3_r_0_rho_1_t_100000}
}

\subfigure[]{
\includegraphics[width=0.31\textwidth]{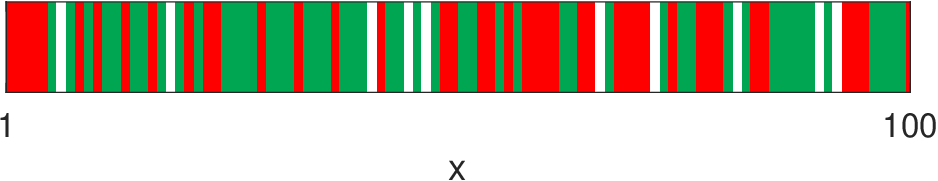}
\label{fig:pattern_abm_p_0_q_0_r_0.9_rho_1_t_0}
}
\subfigure[]{
\includegraphics[width=0.31\textwidth]{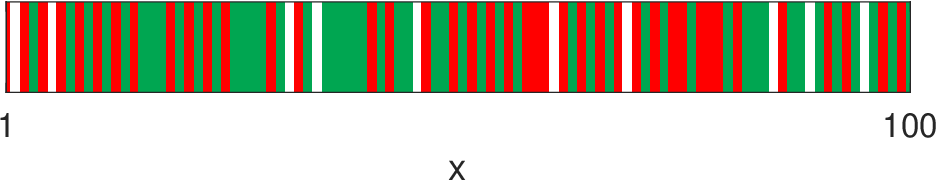}
\label{fig:pattern_abm_p_0_q_0_r_0.9_rho_1_t_1000}
}
\subfigure[]{
\includegraphics[width=0.31\textwidth]{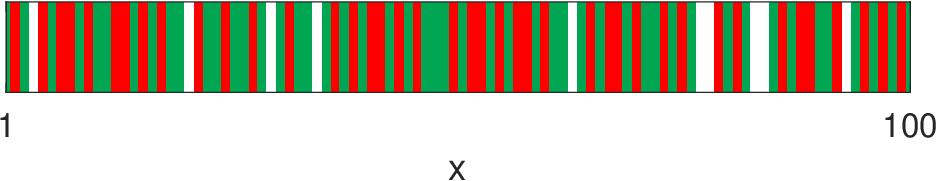}
\label{fig:pattern_abm_p_0_q_0_r_0.9_rho_1_t_100000}
}
\end{center}
\caption{Pattern formation in a single realisation of the ABM. The one-dimensional lattice with $L=100$ sites is initialised with agents of type A and type B. In \subref{fig:pattern_abm_p_0.6_q_0.55_r_0.5_rho_1_t_0} species A and species B are initially segregated. Each site on the left half of the lattice (corresponding to $1\leqslant i \leqslant 50$) is seeded with a type-A agent with probability $\pi_1=0.9$ or the site is left empty otherwise. Each site on the right half of the lattice (corresponding to $51\leqslant i \leqslant 100$) is seeded with a type-B agent with probability $\pi_2=0.9$ or left empty otherwise. In \subref{fig:pattern_abm_p_0.9_q_0.3_r_0_rho_1_t_0} and \subref{fig:pattern_abm_p_0_q_0_r_0.9_rho_1_t_0} each site is occupied with a type-A agent with probability $\pi_1=0.45$ or a type-B agent with probability $\pi_2=0.45$. A site is left empty with probability $1-\pi_1-\pi_2$. We let the positions of the agents evolve according to the two-species ABM and present the densities of species A and B at $t=$[0, 1,000, 100,000]. The three simulations evolve to three distinct types of patterns: mixing for $p=0.6,\, q=0.55, \, r=0.5$ in panels \subref{fig:pattern_abm_p_0.6_q_0.55_r_0.5_rho_1_t_0}-\subref{fig:pattern_abm_p_0.6_q_0.55_r_0.5_rho_1_t_100000}, aggregation for $p=0.9,\, q=0.3, \, r=0$ in panels \subref{fig:one_dimensional_ts_sme_p_0.9_q_0.3_r_0_rho_1_t_0}-\subref{fig:one_dimensional_ts_sme_p_0.9_q_0.3_r_0_rho_1_t_100000} and one-dimensional checkerboard (i.e. alternating) for $p=0,\, q=0, \, r=0.9$ in panels \subref{fig:one_dimensional_ts_sme_p_0_q_0_r_0.9_rho_1_t_0}-\subref{fig:one_dimensional_ts_sme_p_0_q_0_r_0.9_rho_1_t_100000}. In all cases, we assume the rate of movements $m_A=m_B=1$ and $\rho=1$ with reflecting boundaries.}
\label{fig:abm_patterns}
\end{figure}

In Figure \ref{fig:ts_1D_abm_sme_comparison} we compare the average behaviour of the ABM with the numerical solution of the SMEs for the same parameters as Figure \ref{fig:ts_1D_abm_pde_comparison_bad}\subref{fig:one_dimensional_ts_p_0.5_q_0_r_0.25_rho_0.5}-\subref{fig:one_dimensional_ts_p_0_q_0.5_r_0_rho_1} (for which we saw poor agreement between the ABM and the PDE previously). We can immediately see that the agreement between the ABM and SMEs is better than the PDEs as the SMEs more closely resemble the average behaviour of the ABM.

\begin{figure}[ht!]
\begin{center}

\subfigure[]{
\includegraphics[width=0.31\textwidth]{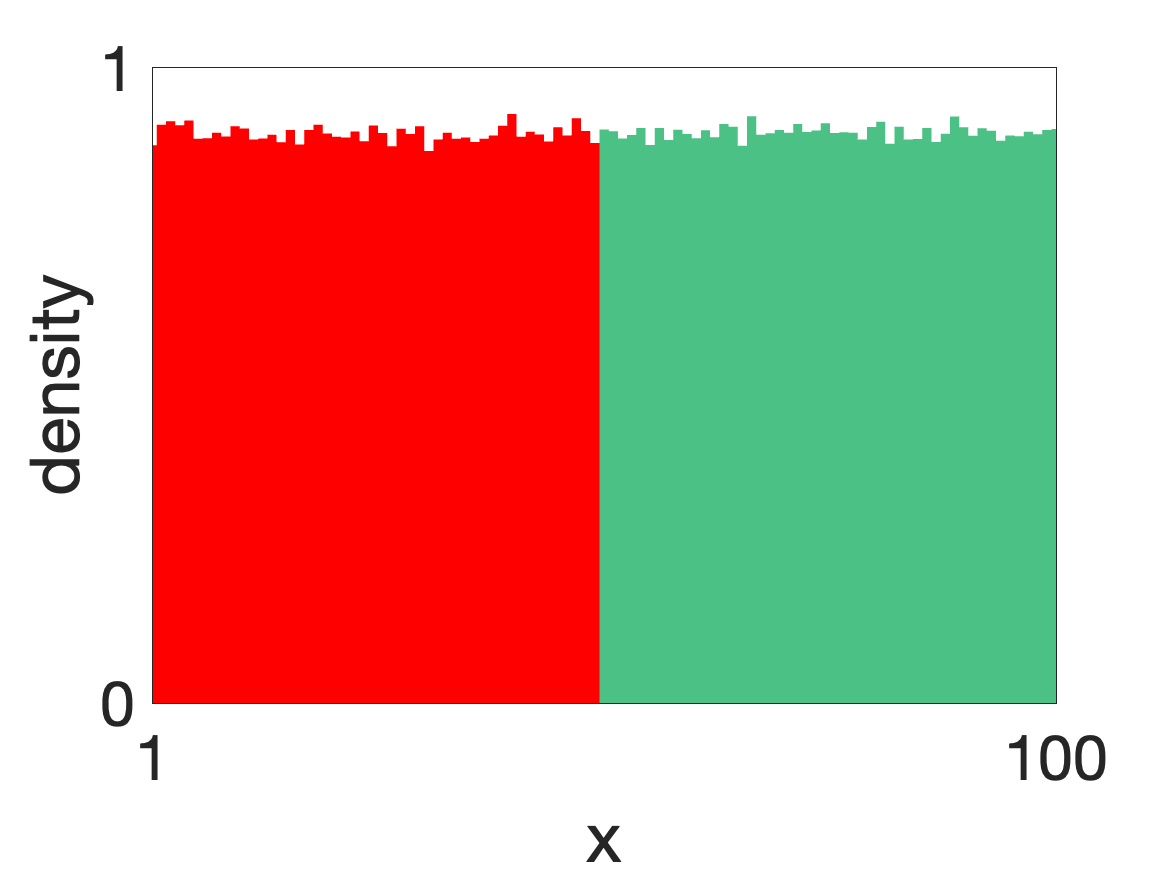}
\label{fig:one_dimensional_ts_sme_p_0.6_q_0.55_r_0.5_rho_1_t_0}
}
\subfigure[]{
\includegraphics[width=0.31\textwidth]{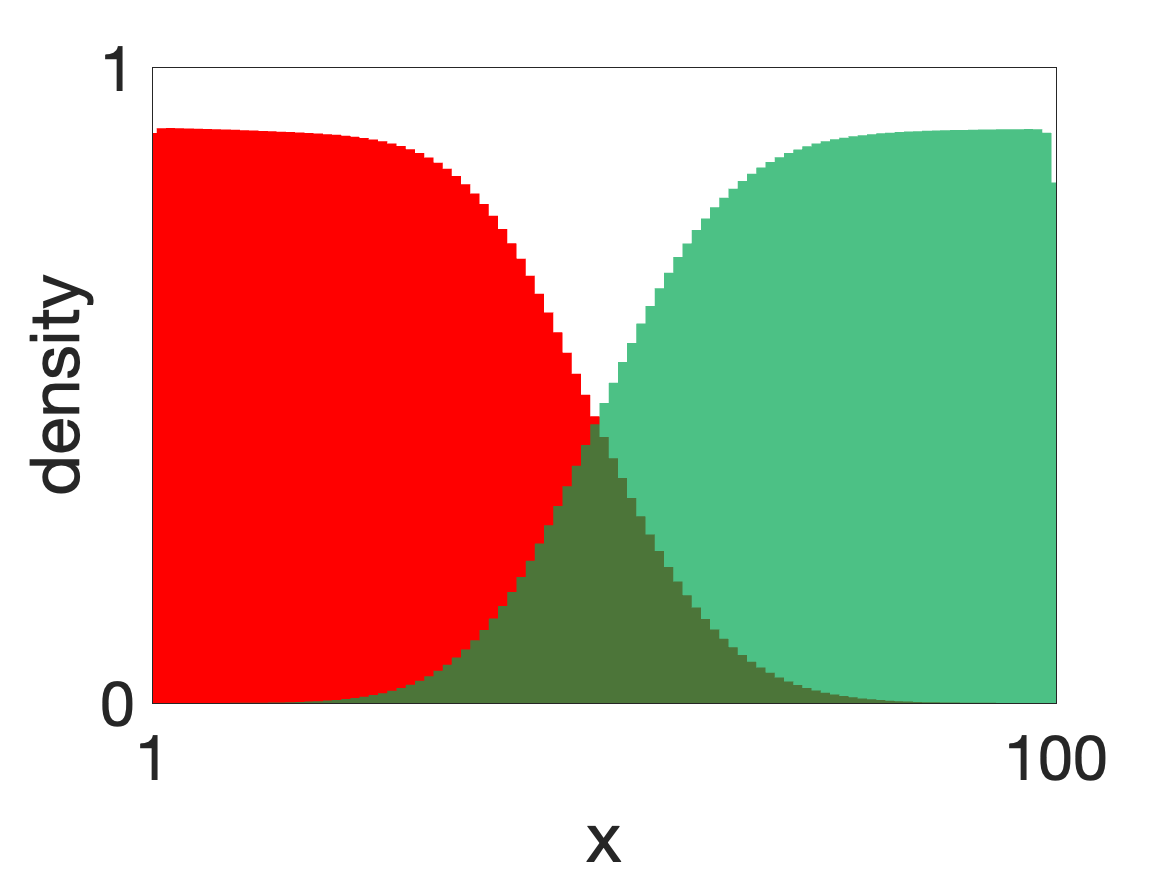}
\label{fig:one_dimensional_ts_sme_p_0.6_q_0.55_r_0.5_rho_1_t_1000}
}
\subfigure[]{
\includegraphics[width=0.31\textwidth]{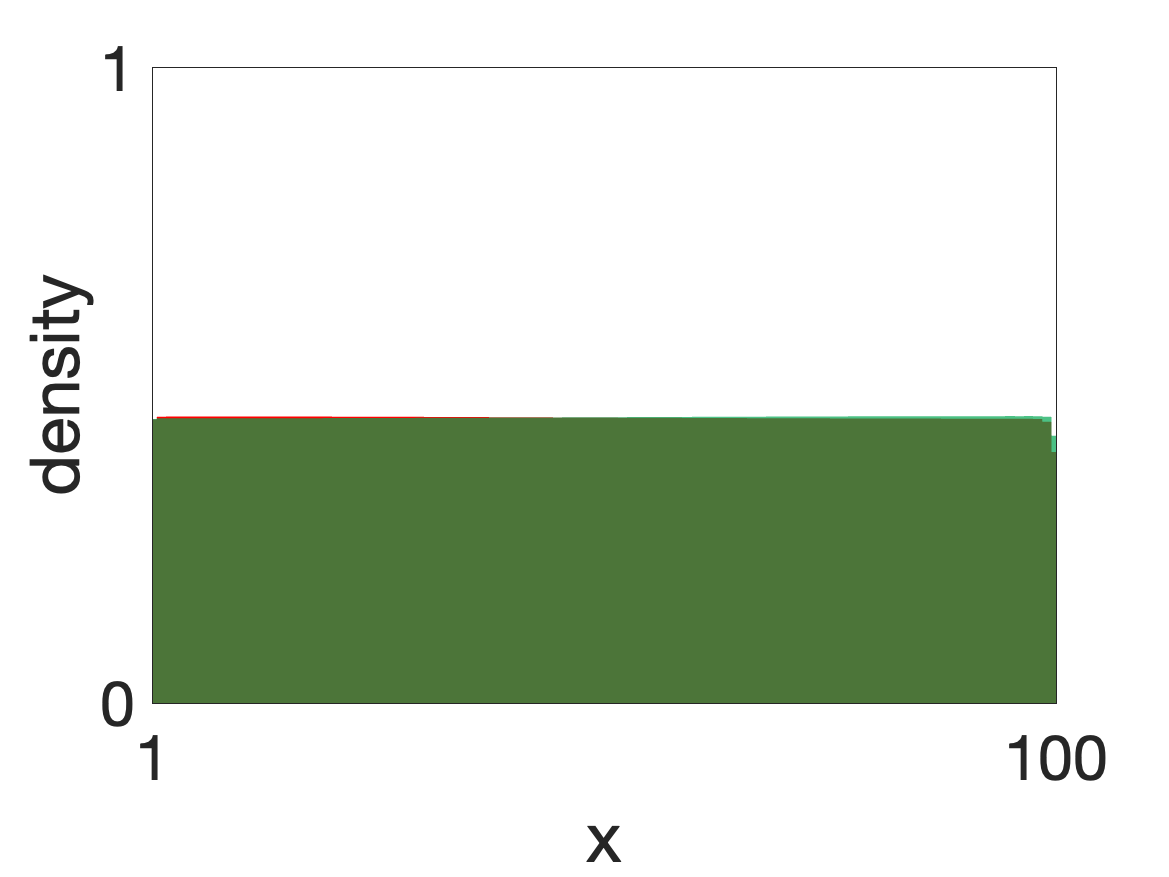}
\label{fig:one_dimensional_ts_sme_p_0.6_q_0.55_r_0.5_rho_1_t_100000}
}

\subfigure[]{
\includegraphics[width=0.31\textwidth]{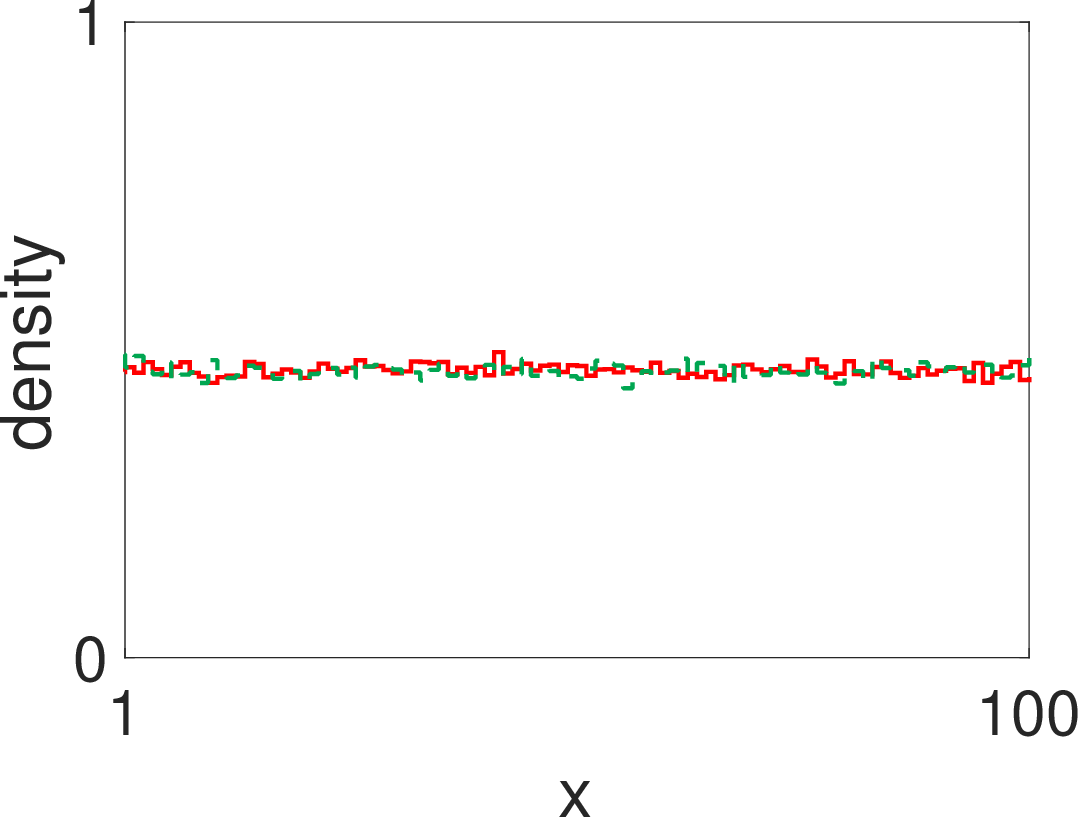}
\label{fig:one_dimensional_ts_sme_p_0.9_q_0.3_r_0_rho_1_t_0}
}
\subfigure[]{
\includegraphics[width=0.31\textwidth]{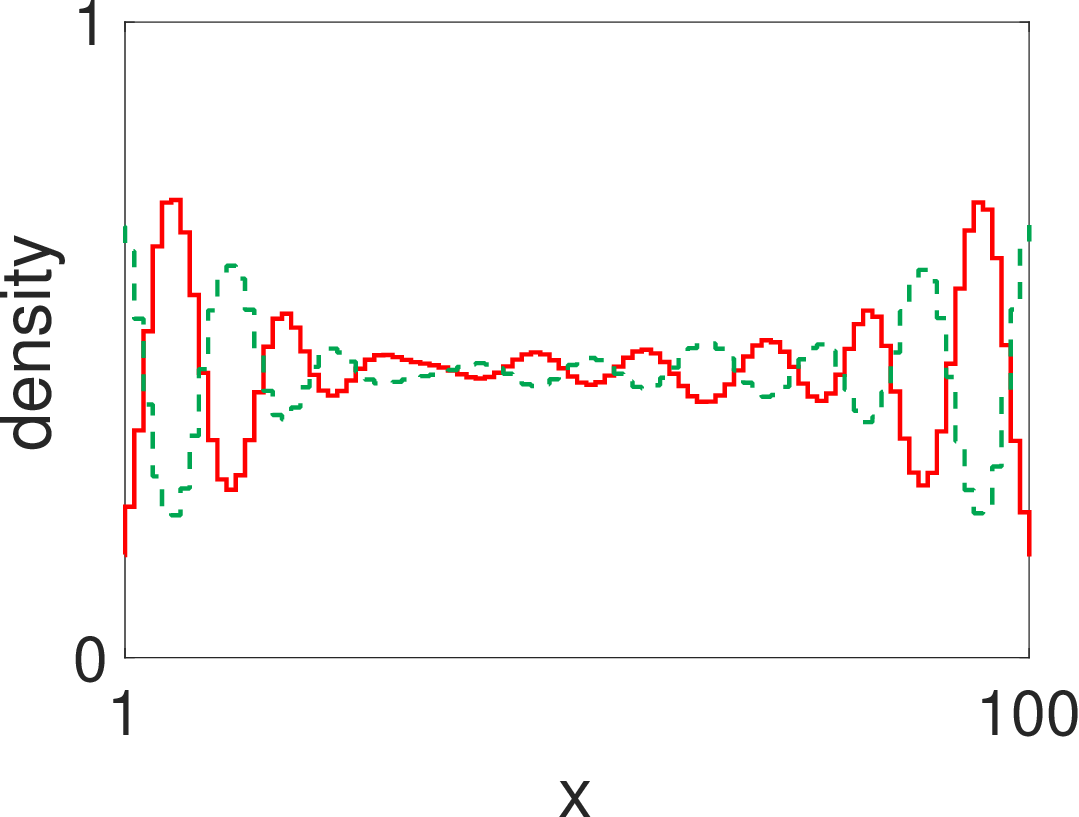}
\label{fig:one_dimensional_ts_sme_p_0.9_q_0.3_r_0_rho_1_t_1000}
}
\subfigure[]{
\includegraphics[width=0.31\textwidth]{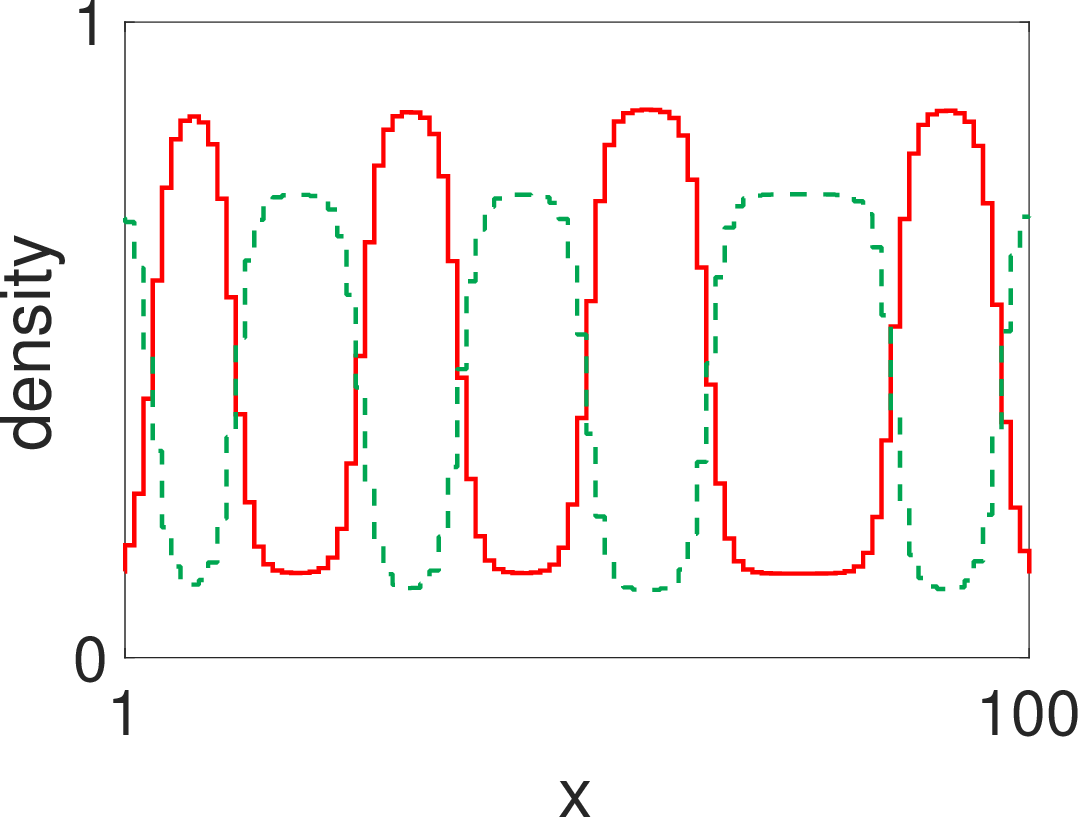}
\label{fig:one_dimensional_ts_sme_p_0.9_q_0.3_r_0_rho_1_t_100000}
}

\subfigure[]{
\includegraphics[width=0.31\textwidth]{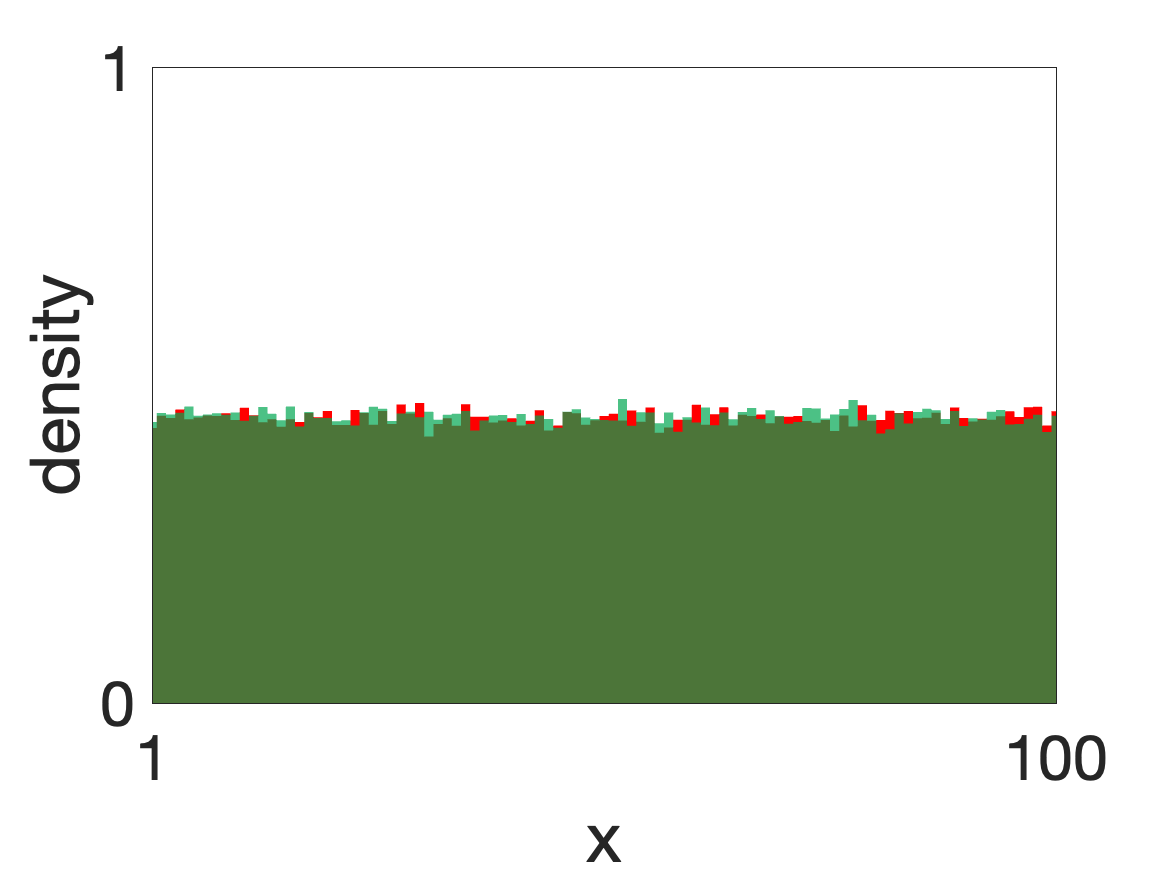}
\label{fig:one_dimensional_ts_sme_p_0_q_0_r_0.9_rho_1_t_0}
}
\subfigure[]{
\includegraphics[width=0.31\textwidth]{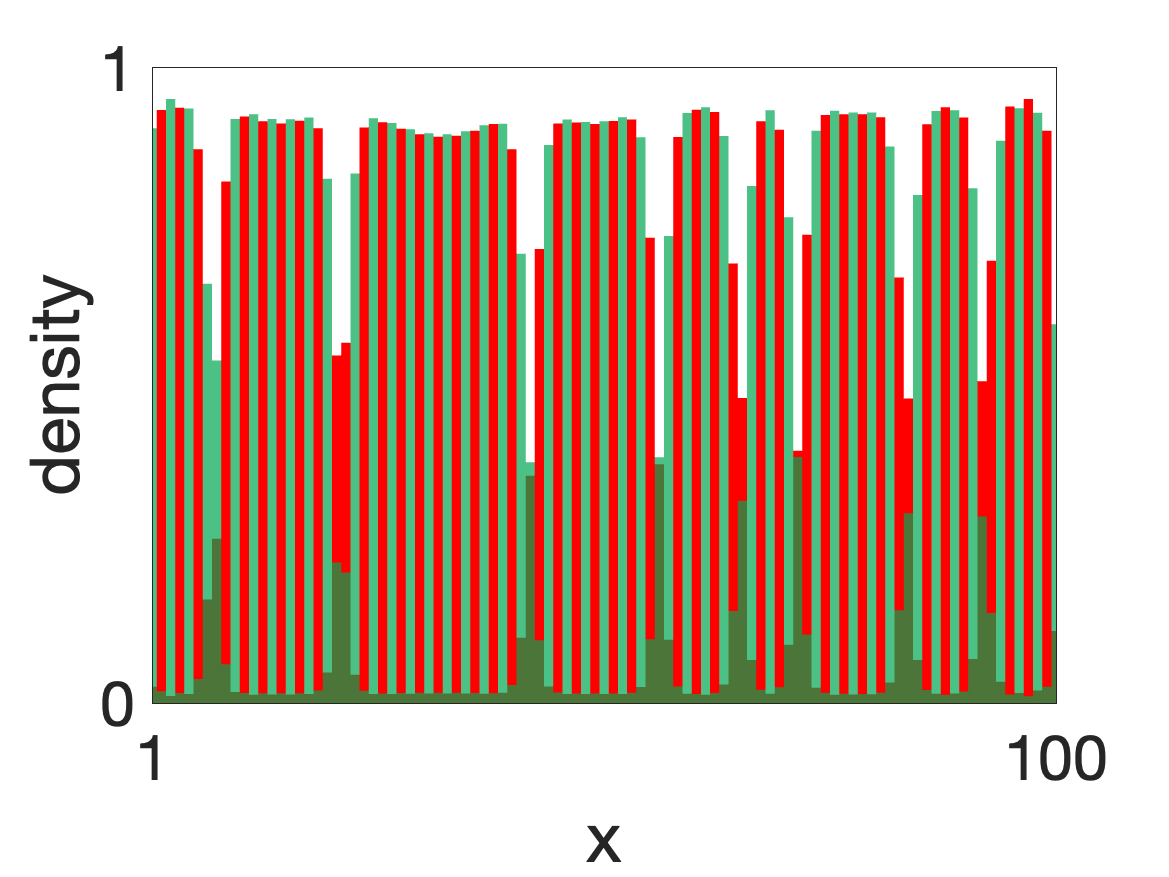}
\label{fig:one_dimensional_ts_sme_p_0_q_0_r_0.9_rho_1_t_1000}
}
\subfigure[]{
\includegraphics[width=0.31\textwidth]{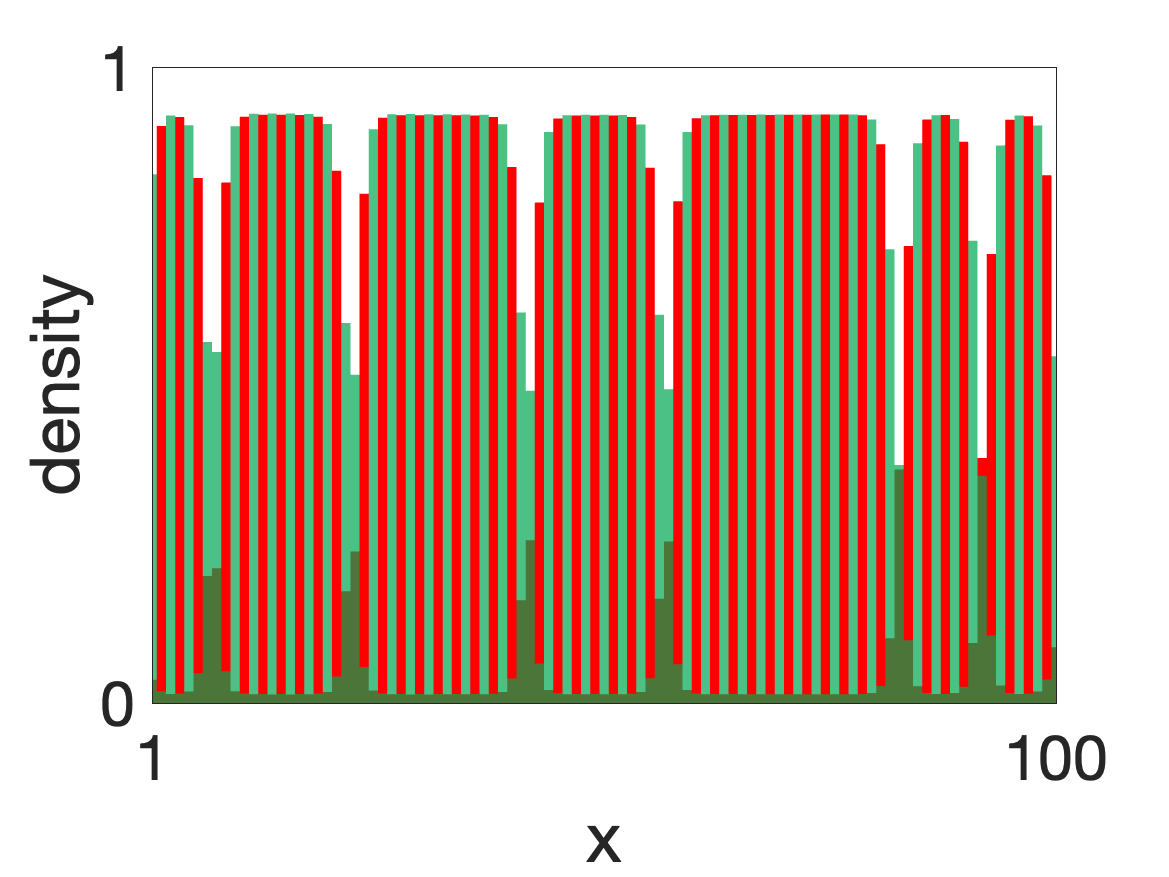}
\label{fig:one_dimensional_ts_sme_p_0_q_0_r_0.9_rho_1_t_100000}
}

\end{center}
\caption{Pattern formation in the SMEs. The one-dimensional domain with $L=100$ sites is initialised with mean densities of species A and B. In \subref{fig:pattern_abm_p_0.6_q_0.55_r_0.5_rho_1_t_0} species A and species B are initially segregated where each site on the left half of the lattice (corresponding to $1\leqslant i \leqslant 50$) is given a mean density $c_A=0.9$ of species A and each site on the right half of the lattice (corresponding to $51\leqslant i \leqslant 100$) is assigned a mean density $c_B=0.9$ of species B. In \subref{fig:pattern_abm_p_0.9_q_0.3_r_0_rho_1_t_0} and \subref{fig:pattern_abm_p_0_q_0_r_0.9_rho_1_t_0} each lattice site is assigned a mean density of $c_A=0.45$ and $c_B=0.45$. In all three scenarios, the initial density was perturbed by adding small independent normally distributed random variables (with mean 0 and variance $0.01$) to the mean density of each species at each lattice site (for symmetry breaking). The SMEs were solved numerically with zero-flux boundary conditions and densities of species A and B at $t=$[0, 1,000, 100,000] showing evolution to three types of patterns are presented here: mixing for $p=0.6,\, q=0.55, \, r=0.5$ in panels \subref{fig:pattern_abm_p_0.6_q_0.55_r_0.5_rho_1_t_0}-\subref{fig:pattern_abm_p_0.6_q_0.55_r_0.5_rho_1_t_100000}, aggregation for $p=0.9,\, q=0.3, \, r=0$ in panels \subref{fig:one_dimensional_ts_sme_p_0.9_q_0.3_r_0_rho_1_t_0}-\subref{fig:one_dimensional_ts_sme_p_0.9_q_0.3_r_0_rho_1_t_100000} and checkerboard for $p=0,\, q=0, \, r=0.9$ in panels \subref{fig:one_dimensional_ts_sme_p_0_q_0_r_0.9_rho_1_t_0}-\subref{fig:one_dimensional_ts_sme_p_0_q_0_r_0.9_rho_1_t_100000}\textsuperscript{1}. In all cases, we assume the rate of movements $m_A=m_B=1$ and $\rho=1$.}
\label{fig:sme_patterns}
\end{figure}

To explore whether our ABM and SMEs are capable of replicating the sorting behaviour of biological cells, we investigate the ABM and the SMEs with different sets of adhesion parameters. We let the positions of the agents evolve according to the ABM and solve the corresponding SMEs numerically with reflecting boundaries. In Figures \ref{fig:abm_patterns} and \ref{fig:sme_patterns} we exemplify three types of behaviours shown by our model in the ABM and the SME (initial data and parameters in figure caption). In panels \subref{fig:one_dimensional_ts_sme_p_0.6_q_0.55_r_0.5_rho_1_t_0}-\subref{fig:one_dimensional_ts_sme_p_0.6_q_0.55_r_0.5_rho_1_t_100000} subject to a mixture of high to moderate adhesion strengths ($p=0.6, \; q=0.55, \; r=0.5$) we see mixing of the two populations to a uniform steady state. More complex patterns are observed in panels \subref{fig:one_dimensional_ts_sme_p_0.9_q_0.3_r_0_rho_1_t_0}-\subref{fig:one_dimensional_ts_sme_p_0.9_q_0.3_r_0_rho_1_t_100000}, in which high self-adhesion between species A and low self-adhesion between species B gives rise to out-of-phase peaks and troughs of density characterising aggregation. In panels \subref{fig:one_dimensional_ts_sme_p_0_q_0_r_0.9_rho_1_t_0}-\subref{fig:one_dimensional_ts_sme_p_0_q_0_r_0.9_rho_1_t_100000} high cross adhesion ($r=0.90$) shows cell sorting behaviour in which the density of lattice sites alternates between species A and B resembling a one-dimensional analogue of a checkerboard pattern. In all three cases we have used a swapping probability of 1 since we found that varying the swapping probability did not fundamentally change the pattern in the SMEs and a high swapping probability led to faster convergence to a pattern.

\footnotetext[1]{In all the panels red denotes the mean density of species A and light green denotes that of species B. In panels \subref{fig:pattern_abm_p_0.6_q_0.55_r_0.5_rho_1_t_0}-\subref{fig:pattern_abm_p_0.6_q_0.55_r_0.5_rho_1_t_100000} and \subref{fig:pattern_abm_p_0_q_0_r_0.9_rho_1_t_0}-\subref{fig:pattern_abm_p_0_q_0_r_0.9_rho_1_t_100000} dark green represents the overlap between the mean densities of the two species.}

\subsection{Two-dimensional model}\label{sec:2D_model}

In the two-dimensional ABM the agents move on a two-dimensional grid that has $L_x$ sites in the horizontal direction and $L_y$ sites in the vertical direction. Naturally, agents on the two-dimensional grid have more neighbours (up, down, left and right) than on a one-dimensional lattice (left and right) which in turn reduces the probability of an agent breaking free from their neighbours due to more adhesive forces from the additional neighbours.

\begin{figure}[b!]
\begin{center}

\subfigure[]{
\includegraphics[width=0.30\textwidth]{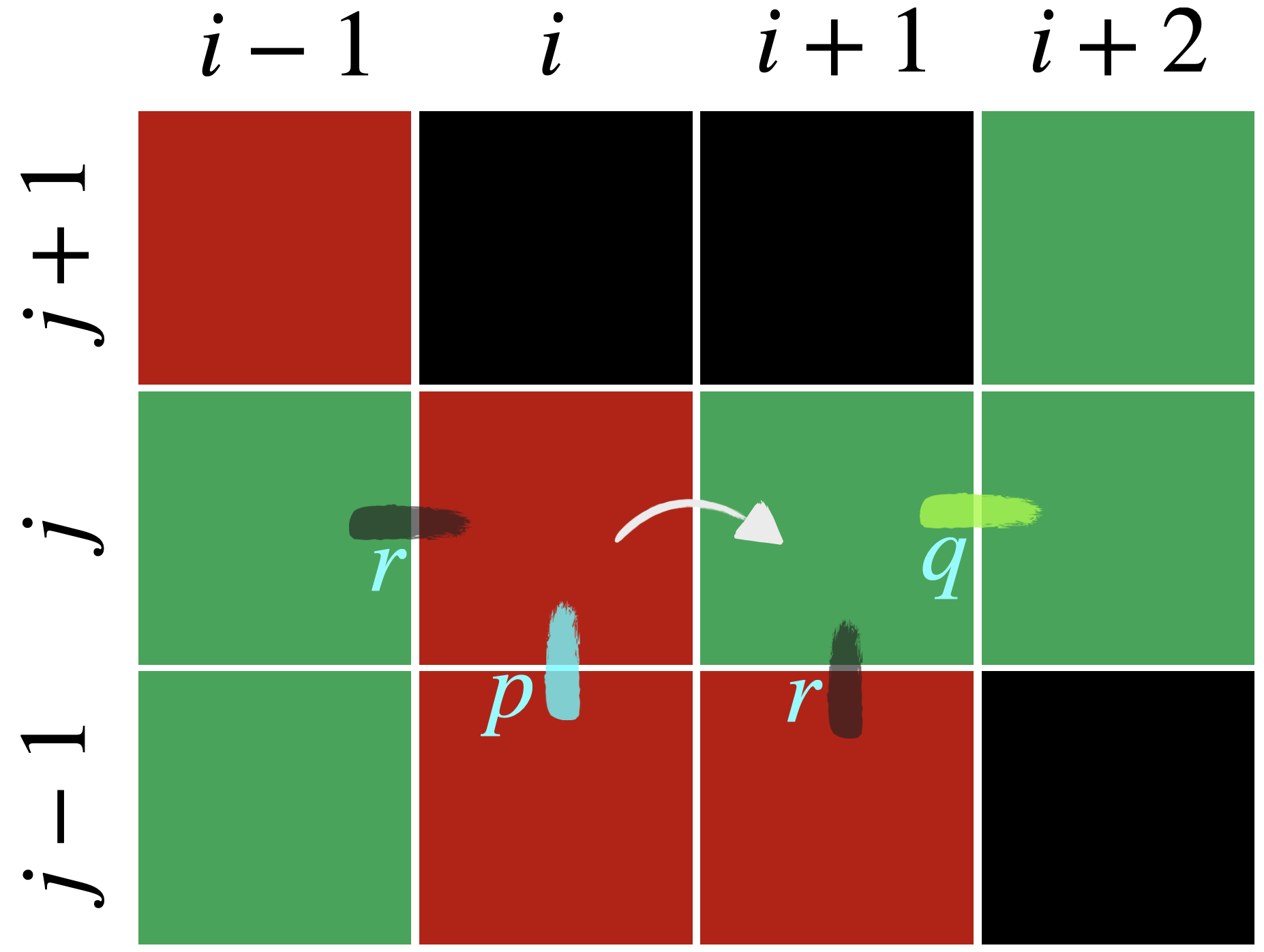}
\label{fig:2d_lattice_1}
}
\subfigure[]{
\includegraphics[width=0.30\textwidth]{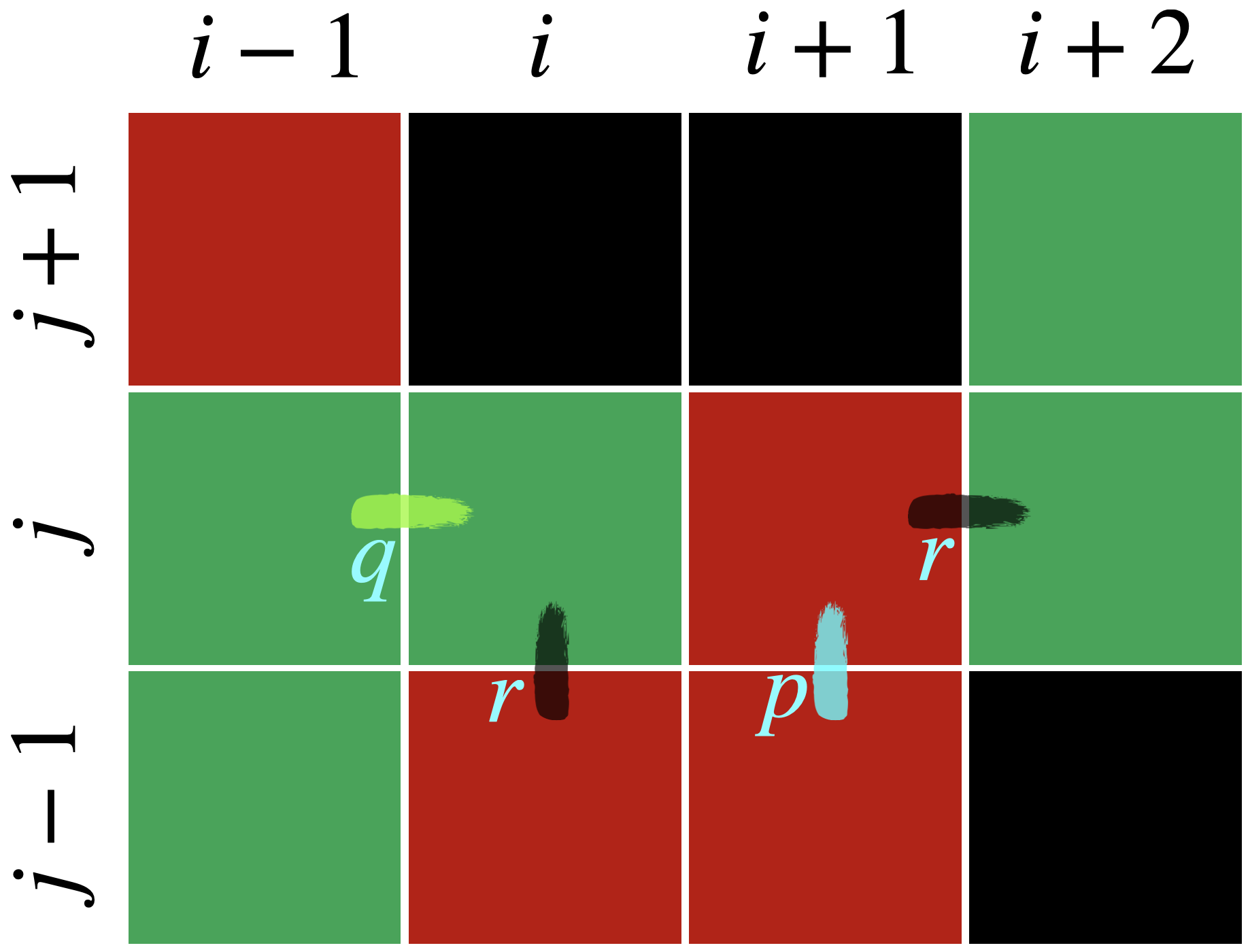}
\label{fig:2d_lattice_2}
}
\end{center}
\caption{A schematic illustrating an example of the two-dimensional ABM with adhesion. A red site is occupied by a type-A agent, a green site represents a type-B agent and a black site is empty. Adhesion strengths between neighbouring pairs of agents are shown as letters $p$, $q$ and $r$ alongside a dash representing a bond between a pair of agents. For clarity, we have only shown labels for three pairs but adhesion is assumed between all neighbouring pairs of agents. The initial configuration of the lattice is shown in panel \subref{fig:2d_lattice_1}. The arrow symbolises the red agent being chosen to move to a site to the right. In panel \subref{fig:2d_lattice_2} we show the result of a successful swap. Note the labels denoting the adhesive bond between pairs of neighbouring agents have also been updated.}
\label{fig:lattice_2D}
\end{figure}

In Figure \ref{fig:lattice_2D} we show an example of an agent swapping into an already occupied site to the right of it on a two-dimensional lattice. The moving agent is of type A (red) with one type-A agent and two type-B agents (green) in its neighbourhood, thus the probability of it breaking adhesive bonds with its neighbours is $(1-p)(1-r)^2$. The agent at the target site is a type-B agent with 1 type-B agent and two type-A agents in its neighbourhood and therefore the probability of the agent at the target site breaking adhesive bonds with its neighbours is $(1-q)(1-r)^2$. Thus the probability that the swap is successful is $\rho (1-p)(1-q)(1-r)^4$. In the case of a successful swap the two agents make new connections with their new neighbours (see panel \subref{fig:2d_lattice_2}).

The PDEs for the densities of species A and species B in the two-dimensional setting can be derived from the corresponding two-dimensional master equations in a similar way to the one-dimensional PDEs. The two-dimensional master equations for species A and B can be found in Appendix \ref{appendix_A}.

The continuum PDEs describing the evolution of the densities of species A and species B are given by,

\begin{align}\label{eqn:ts_2D_pde_A}
& \D A t = \nabla \cdot \Bigg[D_A\Big(D_1^\text{2D}(A,B) \nabla B +D_2^\text{2D}(A,B)\nabla A \Big)
\nonumber\\
& \quad + (D_A+D_B)\rho \Big(D_3^\text{2D}(A,B)\nabla A - D_4^\text{2D}(A,B)\nabla B \Big)\Bigg],
\end{align}

\begin{align}\label{eqn:ts_2D_pde_B}
& \D B t = \nabla \cdot \Bigg[D_B \Big(D_5^\text{2D}(A,B) \nabla A + D_6^\text{2D}(A,B) \nabla B \Big) \nonumber\\ & \quad + (D_A+D_B)\rho \Big(D_7^\text{2D}(A,B) \nabla B - D_8^\text{2D}(A,B) \nabla A \Big)\Bigg],
\end{align}

where,

\begin{equation}\label{eqn:diff_constant_2D}
D_A= \lim_{\substack{\Delta \to 0 \\ \delta t \to 0}}\frac{m_A}{4}\frac{\Delta^2}{\delta t}, \quad D_B= \lim_{\substack{\Delta \to 0 \\ \delta t \to 0}}\frac{m_B}{4}\frac{\Delta^2}{\delta t},
\end{equation}

and,

\begin{align*}
D_1^\text{2D}(A,B)&=(1-p)^{4A}(1-r)^{4B}(4 \ln(1-r)(1-A-B)+1)A, \\
D_2^\text{2D}(A,B)&=(1-p)^{4A}(1-r)^{4B}(4\ln(1-p)A(1-A-B)+1-B),\\
D_3^\text{2D}(A,B)&=(1-p)^{4A}(1-q)^{4B}(1-r)^{4A+4B}B(1-4\ln(1-r)A+4\ln(1-p)A),\\
D_4^\text{2D}(A,B)&=(1-p)^{4A}(1-q)^{4B}(1-r)^{4A+4B}A(1-4\ln(1-q)B+4\ln(1-r)B),\\
D_5^\text{2D}(A,B)&=(1-q)^{4B}(1-r)^{4A}(4 \ln(1-r)(1-A-B)+1)B, \\
D_6^\text{2D}(A,B)&=(1-q)^{4B}(1-r)^{4A}(4\ln(1-q)B(1-A-B)+1-A),\\
D_7^\text{2D}(A,B)&=(1-q)^{4B}(1-p)^{4A}(1-r)^{4A+4B}(A(1-4\ln(1-r)B+4\ln(1-q)B),\\
D_8^\text{2D}(A,B)&=(1-q)^{4B}(1-p)^{4A}(1-r)^{4A+4B}B(1-4\ln(1-p)A+4\ln(1-r)A). \numberthis \label{eqn:two_species_diff_coeff}
\end{align*}

\begin{figure}[t!]
\begin{center}
\subfigure[]{
\includegraphics[width=0.31\textwidth]{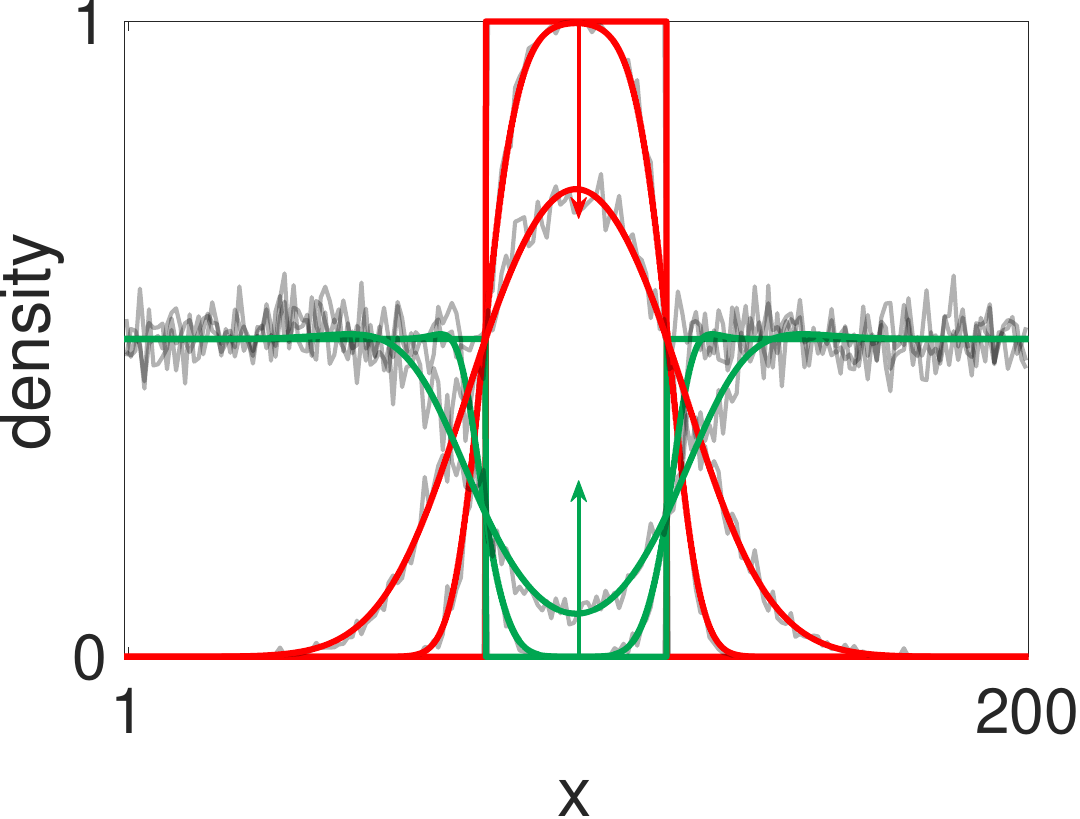}
\label{fig:2D_pde_p_0_q_0.25_r_0_rho_0.25}
}
\subfigure[]{
\includegraphics[width=0.31\textwidth]{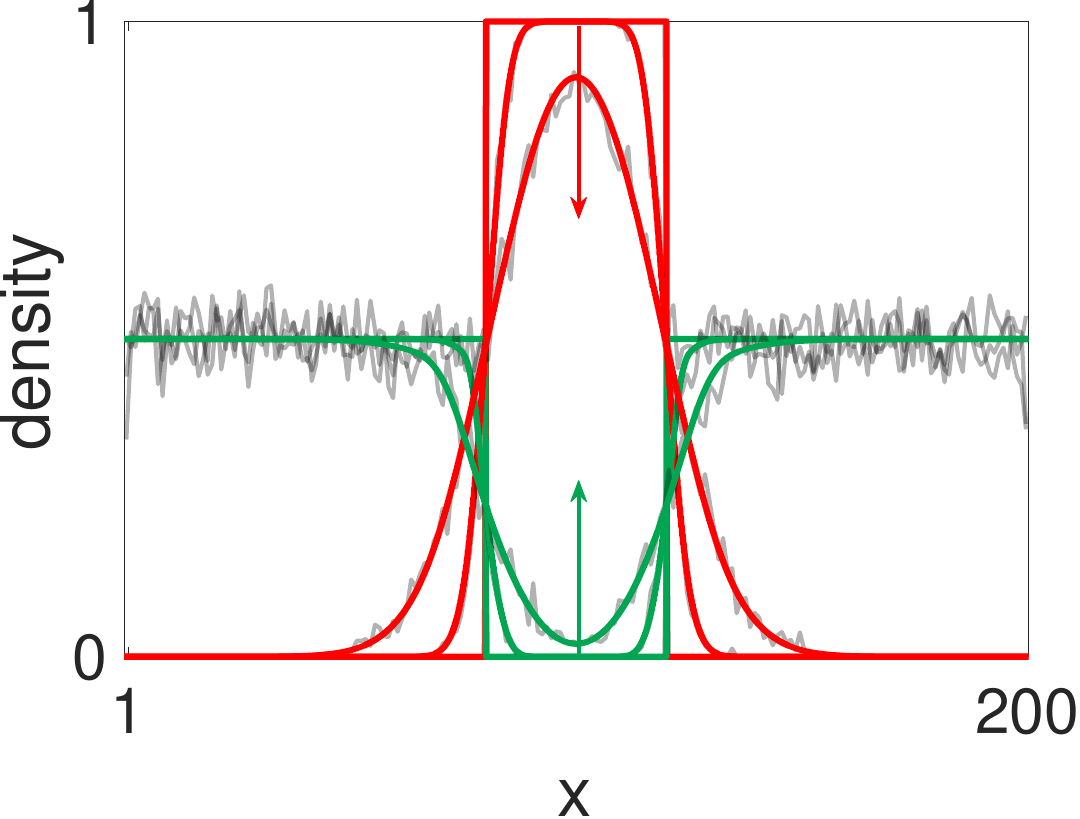}
\label{fig:2D_pde_p_0.25_q_0.5_r_0.25_rho_1}
}
\subfigure[]{
\includegraphics[width=0.31\textwidth]{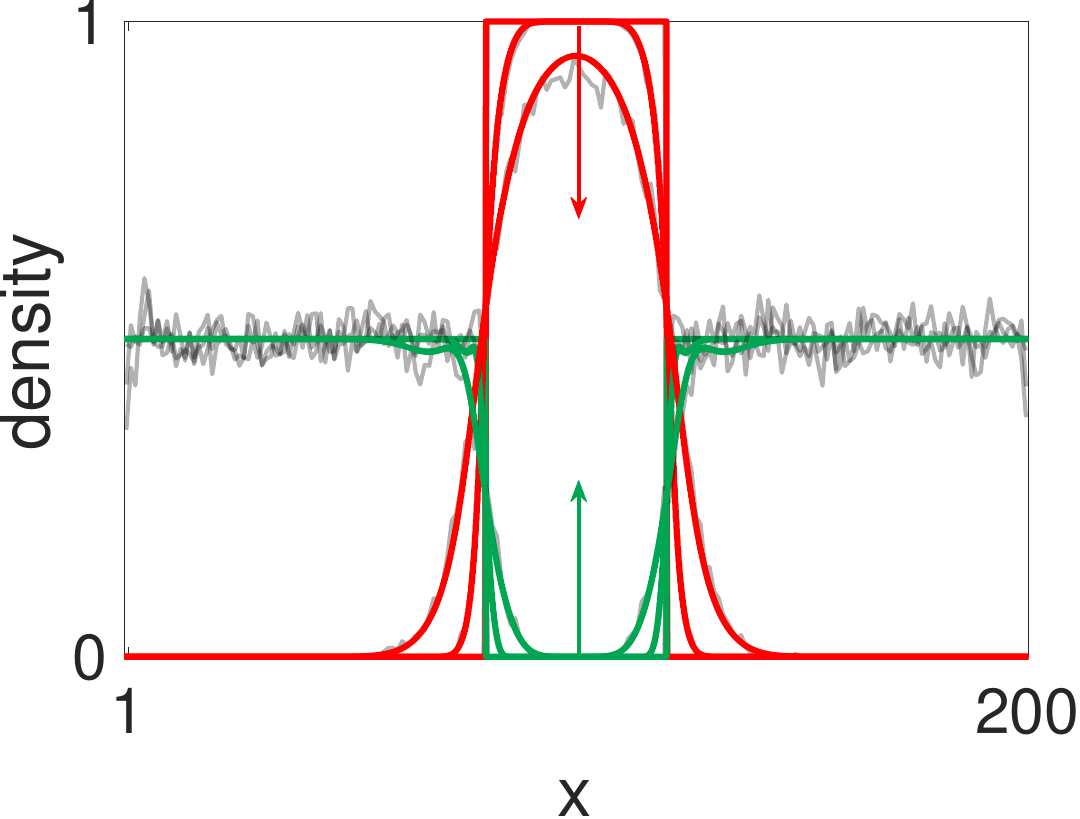}
\label{fig:2D_pde_p_0.25_q_0.5_r_0.5_rho_0.75}
}
\end{center}
\caption{Comparison between the two species column-averaged ABM and the vertically-averaged solution of PDEs \eqref{eqn:ts_2D_pde_A} and \eqref{eqn:ts_2D_pde_B} at $t=0, \, 100, \, 1000$ for $m_A=m_B=1$ and \subref{fig:2D_pde_p_0_q_0.25_r_0_rho_0.25} $(p,q,r,\rho)=(0,0.25,0,0.25)$, \subref{fig:2D_pde_p_0.25_q_0.5_r_0.25_rho_1} $(p,q,r,\rho)=(0.25,0.5,0.25,1)$ and \subref{fig:2D_pde_p_0.25_q_0.5_r_0.5_rho_0.75} $(p,q,r,\rho)=(0.25,0.5,0.5,0.75)$. The lattice has dimensions $L_x=200$ columns and $L_y=20$ rows with zero-flux boundary conditions and is initialised such that all the sites for $81 \leqslant x \leqslant 120$ have density 1 for species A and the remainder of the domain is initialised with species B at density 0.5.}
\label{fig:ts_2D_abm_pde_comparison}
\end{figure}

\begin{figure}[t!]
\begin{center}
\subfigure[]{
\includegraphics[width=0.31\textwidth]{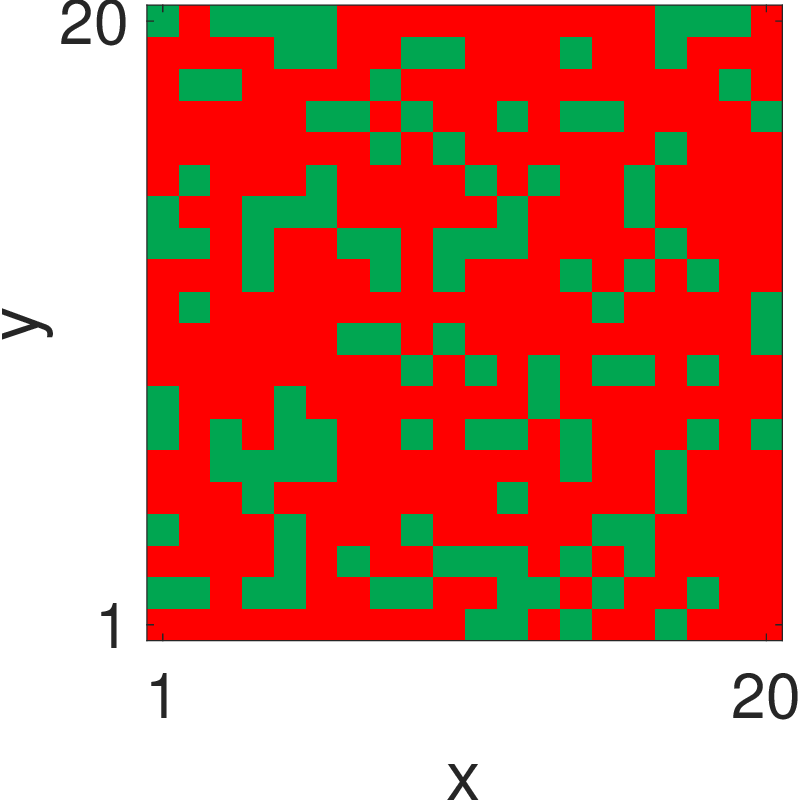}
\label{fig:engulfment_abm_ic}
}
\subfigure[]{
\includegraphics[width=0.31\textwidth]{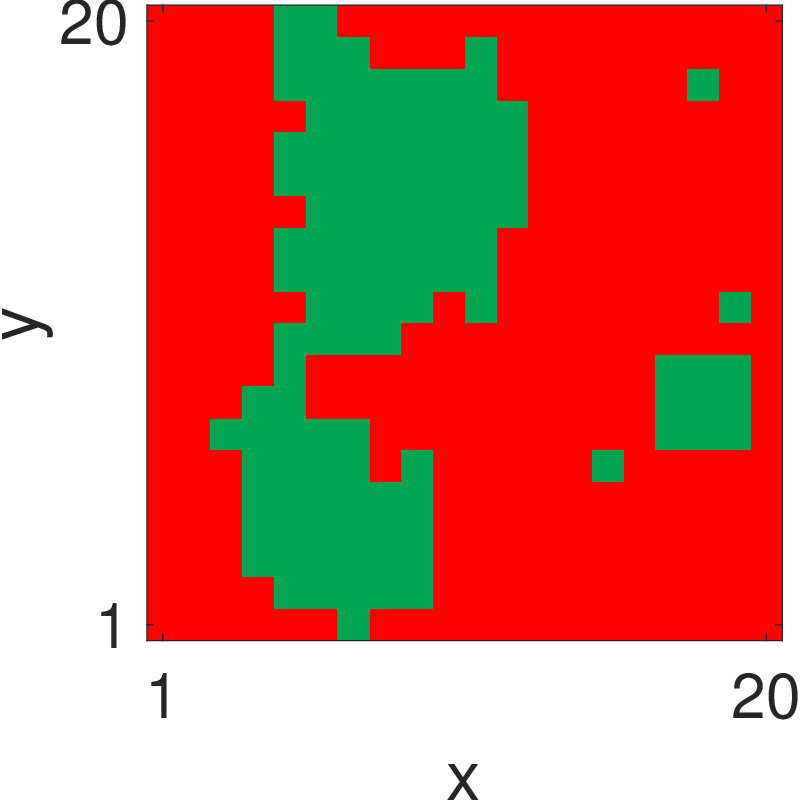}
\label{fig:engulfment_abm_t_10000}
}
\subfigure[]{
\includegraphics[width=0.31\textwidth]{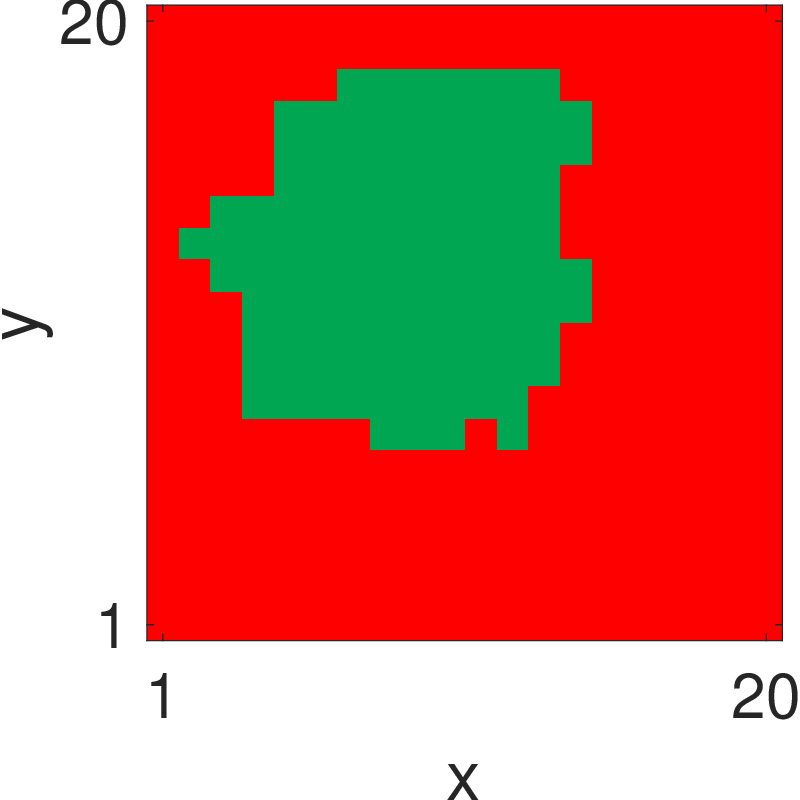}
\label{fig:engulfment_abm_t_100000}
}

\subfigure[]{
\includegraphics[width=0.31\textwidth]{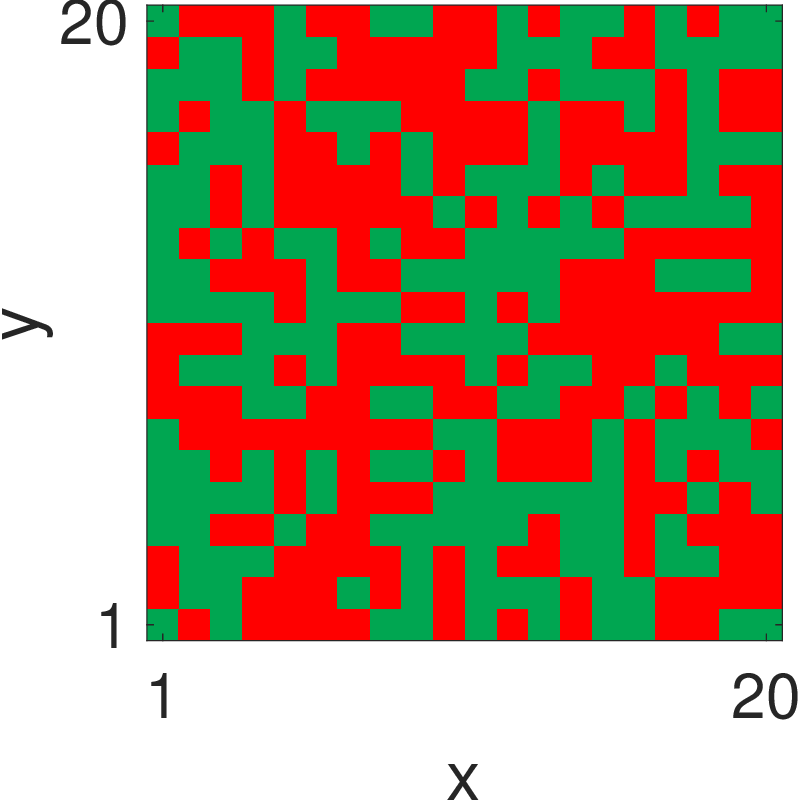}
\label{fig:cheque_abm_ic}
}
\subfigure[]{
\includegraphics[width=0.31\textwidth]{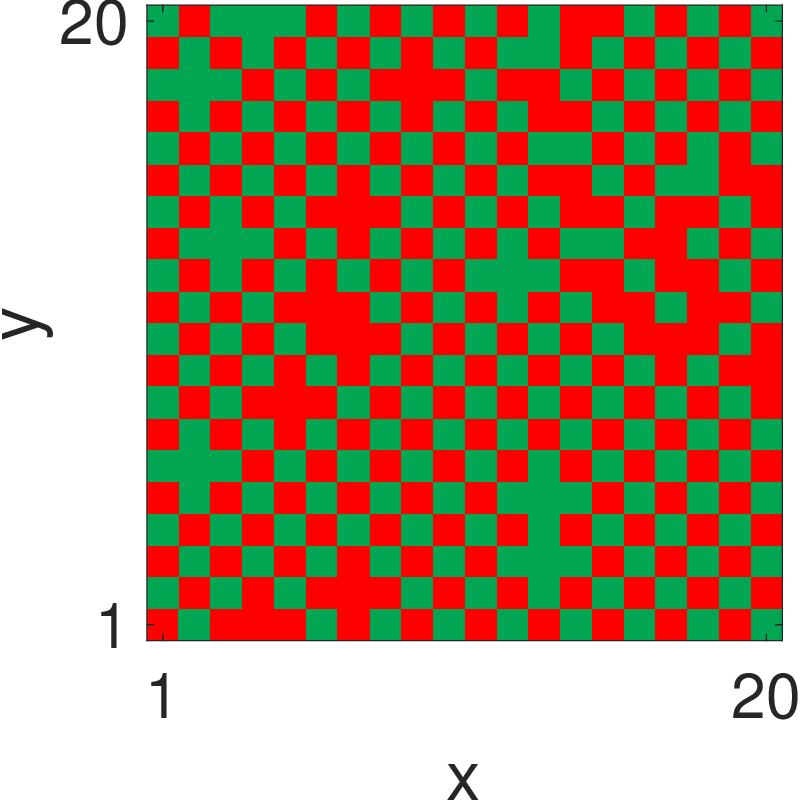}
\label{fig:cheque_abm_t_10000}
}
\subfigure[]{
\includegraphics[width=0.31\textwidth]{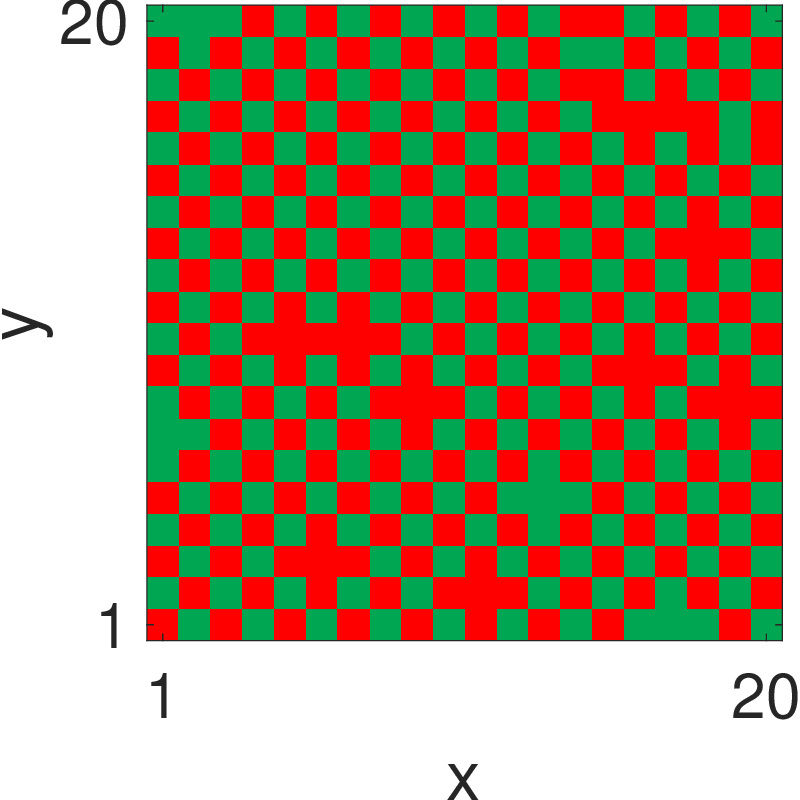}
\label{fig:cheque_abm_t_100000}
}
\end{center}
\caption{Engulfment \subref{fig:engulfment_abm_ic}-\subref{fig:engulfment_abm_t_100000} and checkerboard \subref{fig:cheque_abm_ic}-\subref{fig:cheque_abm_t_100000} patterns in the two-dimensional ABM model on a 20 by 20 lattice from a uniformly random initial condition with reflecting boundaries and $m_A=m_B=\rho=1$. Each lattice site was seeded with a type-A agent with probability $\pi_1$ or a type-B agent with probability $\pi_2=1-\pi_1$. For engulfment in \subref{fig:engulfment_abm_ic} $\pi_1=0.7$ and hence $\pi_2=0.3$ and for the checkerboard simulation in \subref{fig:cheque_abm_ic} $\pi_1=\pi_2=0.5$. We present lattice snapshots at $t=$[0, 10,000, 100,000] for engulfment \subref{fig:engulfment_abm_ic}-\subref{fig:engulfment_abm_t_100000} for $(p,q,r)=(0.2,0.9,0)$ and checkerboard \subref{fig:cheque_abm_ic}-\subref{fig:cheque_abm_t_100000} for $(p,q,r)=(0,0,0.8)$.}
\label{fig:2d_abm_patterns}
\end{figure}

In Figure \ref{fig:ts_2D_abm_pde_comparison} we compare the average column density of the ABM with the vertically-averaged solution of PDEs \eqref{eqn:ts_2D_pde_A} and \eqref{eqn:ts_2D_pde_B} over a lattice with $L_x=200$ and $L_y=20$ with zero-flux boundaries and movement rates $m_A=m_B=1$. Lattice is initialised such that all the sites in the middle 20\% of the domain are fully occupied by type-A agents and the rest of the sites are seeded with type-B agents at density 0.5. For ease of comparison, at $t=0, \, 100$ and 1,000 the densities of the two-dimensional ABM were averaged over the vertical extent of the domain to obtain a one-dimensional representation of the densities of species A and B.

Likewise, the two-dimensional PDEs \eqref{eqn:ts_2D_pde_A} and \eqref{eqn:ts_2D_pde_B} were solved numerically for the densities of both species which were then vertically averaged to obtain one-dimensional analogues. We notice good agreement between the ABM and the PDE for a range of swapping and adhesion parameters. However, we also found that for very strong adhesion, as in the one-dimensional scenario, the PDEs become unstable (figure not shown). Consequently, it is not possible to observe patterns that require strong adhesion in the PDE model. As with the one-dimensional model, we can use the SMEs corresponding to the two-dimensional ABM to analyse our model for pattern formation.

\begin{figure}[t!]
\begin{center}
\subfigure[]{
\includegraphics[width=0.295\textwidth]{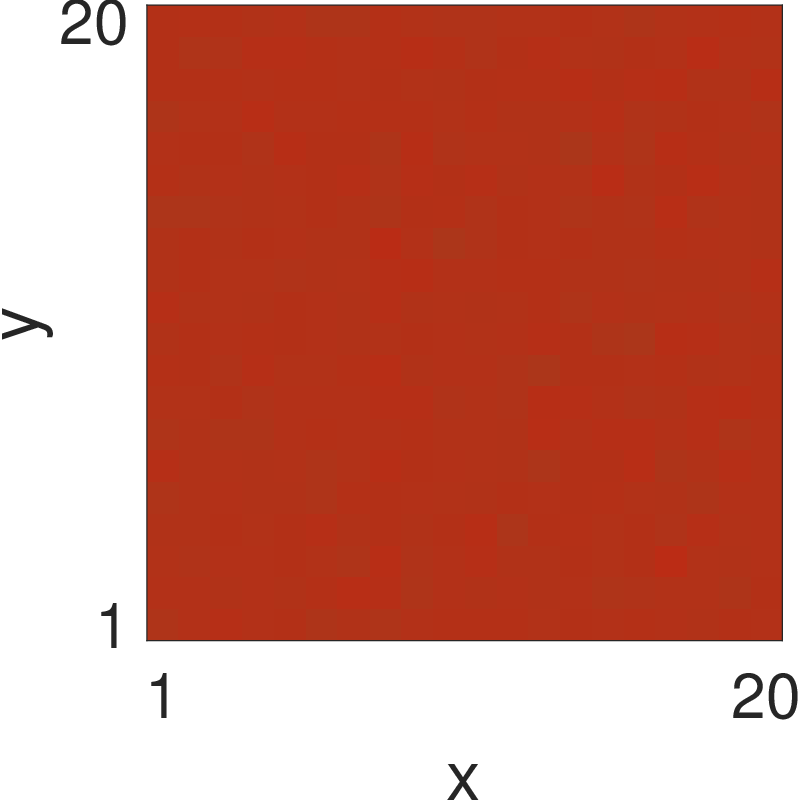}
\label{fig:engulfment_sme_ic}
}
\subfigure[]{
\includegraphics[width=0.295\textwidth]{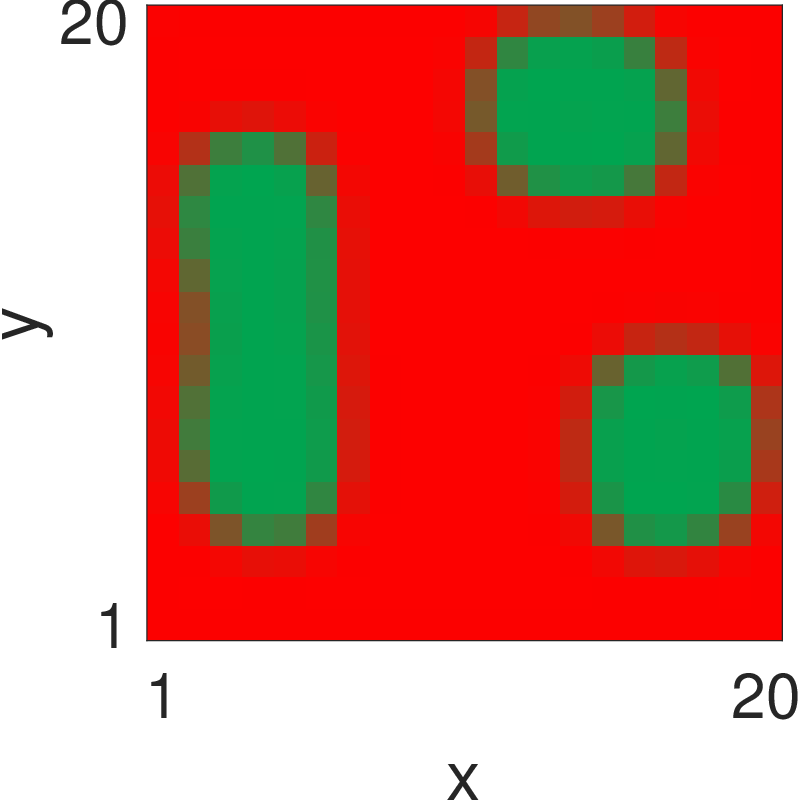}
\label{fig:engulfment_sme_t_10000}
}
\subfigure[]{
\includegraphics[width=0.34\textwidth]{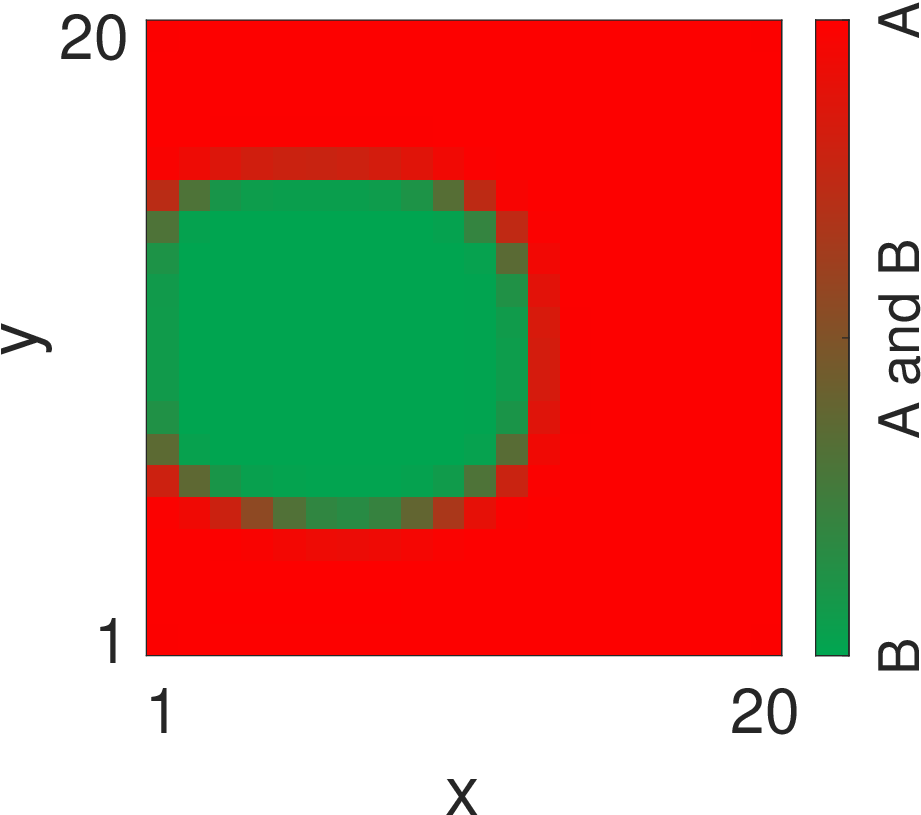}
\label{fig:engulfment_sme_t_100000}
}

\subfigure[]{
\includegraphics[width=0.295\textwidth]{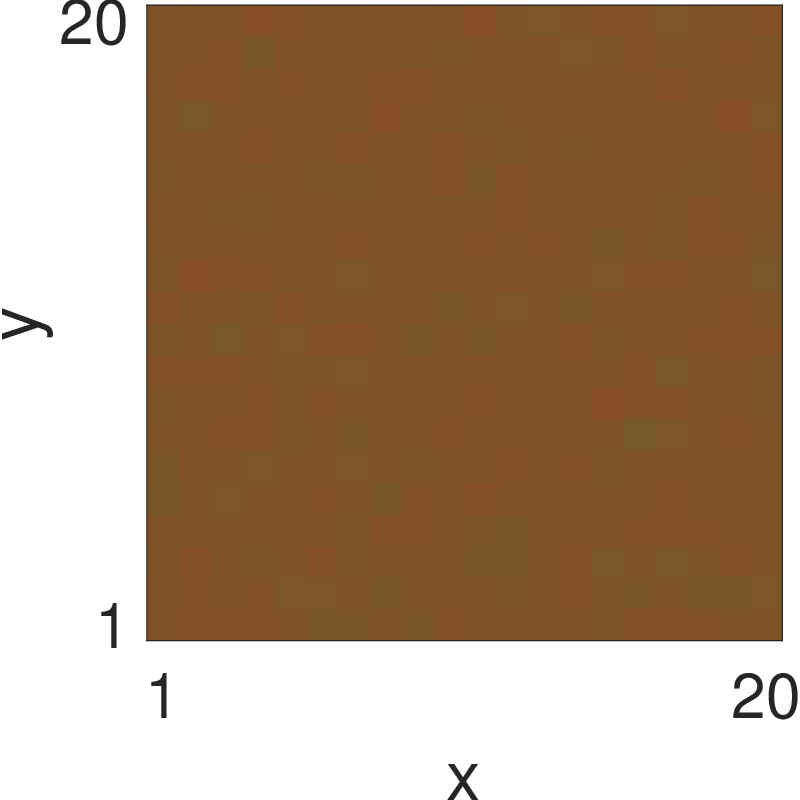}
\label{fig:cheque_sme_ic}
}
\subfigure[]{
\includegraphics[width=0.295\textwidth]{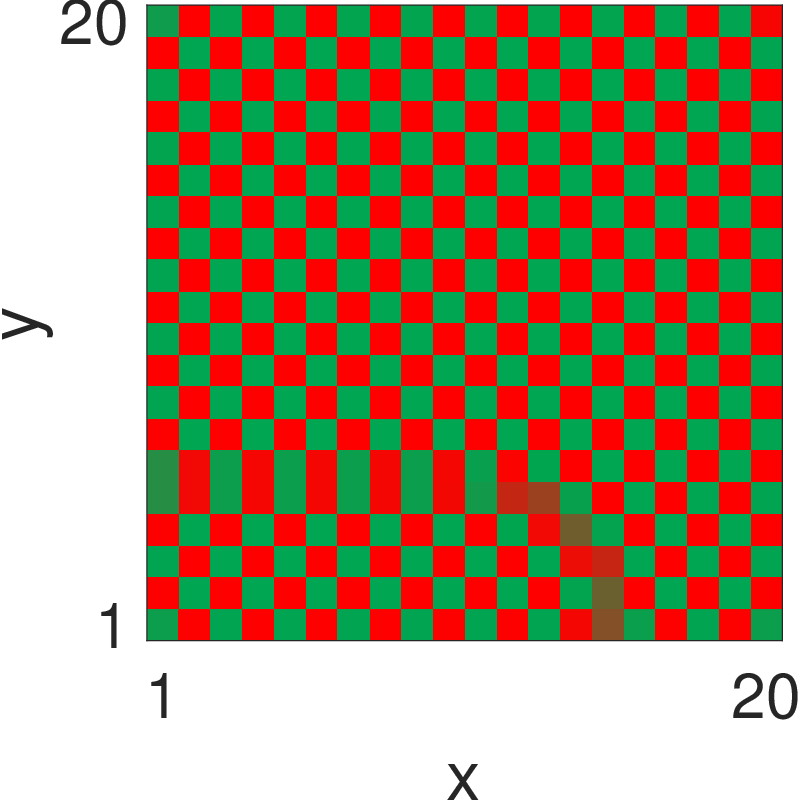}
\label{fig:cheque_sme_t_10000}
}
\subfigure[]{
\includegraphics[width=0.34\textwidth]{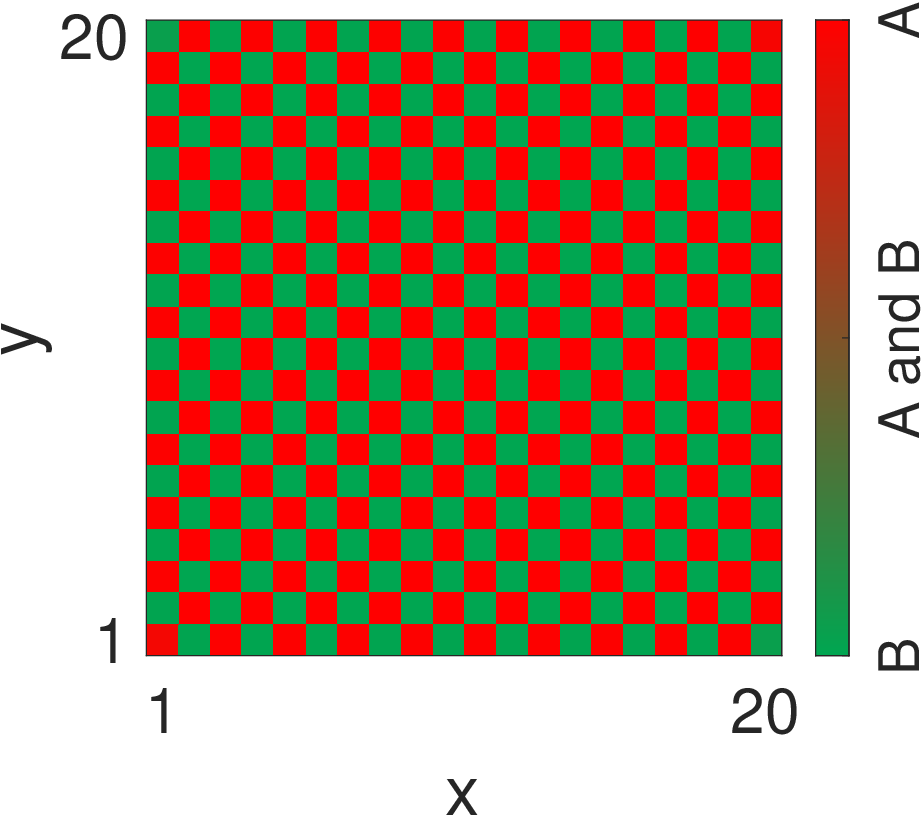}
\label{fig:cheque_sme_t_100000}
}
\end{center}
\caption{Engulfment and checkerboard patterns in the two-dimensional SME model on a 20 by 20 lattice from a uniformly random initial condition with reflecting boundaries with $m_A=m_B=\rho=1$. In both cases, each lattice site was initialised with a constant mean density $c_A$ of species A, which was perturbed by adding independent normally distributed random variables (with zero mean and variance 0.01) to give the perturbed density of species A, $\tilde c_A$. The density of species B at each site was then assigned as $c_B=1-\tilde c_A$. For engulfment, $c_A=0.7$ and for the checkerboard $c_A=0.5$. We present the solution of the SMEs at $t=$[0, 10,000, 100,000] for engulfment \subref{fig:engulfment_sme_ic}-\subref{fig:engulfment_sme_t_100000} for $(p,q,r)=(0.2,0.9,0)$ and the checkerboard \subref{fig:cheque_sme_ic}-\subref{fig:cheque_sme_t_100000} for $(p,q,r)=(0,0,0.8)$. Due to space constraints, the colour bar is exclusively displayed in the rightmost panel for each scenario. Nevertheless, it remains consistent across all panels for that specific scenario.}
\label{fig:2d_sme_patterns}
\end{figure}

Similar to the one-dimensional model, the SMEs can be derived directly from the two-dimensional master equations for species A and B (see Appendix \ref{appendix_B}). In Figures \ref{fig:2d_abm_patterns} and \ref{fig:2d_sme_patterns} we present examples of engulfment and checkerboard patterns obtained with our adhesion model under the ABM and the SMEs, respectively, on a 20 by 20 domain with zero-flux boundaries.

In panels \subref{fig:engulfment_abm_ic}-\subref{fig:engulfment_abm_t_100000} of Figures \ref{fig:2d_abm_patterns} and \ref{fig:2d_sme_patterns} strong self-adhesion between agents of type A, low self-adhesion between agents of type B and no cross-adhesion leads to engulfment of species A by species B over time. Starting from uniformly random densities, species A slowly engulfs species B over time. In panels \subref{fig:cheque_abm_ic}-\subref{fig:cheque_abm_t_100000} high cross-adhesion and no self-adhesion leads to the self-organisation of cells into a checkerboard pattern. The agreement between the ABM and the SME patterns is commendable for the intermediate as well as (and especially) the final pattern. The engulfment and checkerboard patterns seen here are comparable to the type of engulfment observed in the cultures of dissociated chick embryo pigmented retinal epithelial and neural retinal cells in which the neural retinal cells engulf the pigmented retinal epithelial cells \citep{armstrong1971lem} and the checkerboard tiling on the luminal surface of the mature oviduct epithelium in adult Japanese quail \citep{honda1986tpc}.

\section{Conclusion}

In this paper, we have developed a model for cell migration incorporating cell-cell adhesion. By first describing the ABM for the dynamics of individual cells, we derived a set of partial differential equations (PDEs) describing the macroscopic behaviour of the ABM. The work presented here is primarily a response to the lack of multispecies discrete and continuum models with local adhesive interactions mediating contact inhibition and serves to elucidate the proof-of-concept model we have developed. In both, one-dimensional and two-dimensional models, presented in Sections \ref{sec:1D_model} and \ref{sec:2D_model}, respectively, we showed that the microscopic and macroscopic modelling paradigms agree well with each other for some parameter values but do not for others. Particularly, we notice that the PDE descriptions are incapable of replicating adhesion-mediated cell sorting patterns. To overcome this limitation, we derived a set of stochastic mean equations  (SMEs) to represent the mean density of cells as an alternative to the PDEs. We found that not only were the SMEs able to reproduce biologically realistic patterns, but they also showed agreement between the ABM and the SME for certain parameter regimes in which the ABM and the PDEs did not agree well.

We found that our discrete model is capable of replicating biologically realistic patterns such as engulfment and checker-board which are comparable to the engulfment of chick embryo pigmented retinal epithelial cells by neural retinal cells in culture \citep{armstrong1971lem} and the checkerboard pattern in Japanese quail oviduct epithelium eggs \citep{honda1986tpc}, respectively. Our results accord with the differential adhesion hypothesis \citep{steinberg1963rtd}; when two populations with different of adhesive strengths interact, cells with stronger adhesion will prefer to adhere to each other and exclude the species with weaker adhesion as this results in more energetically favourable configurations \citep{dickinson2020tmc}. In the case of the checkerboard pattern (see Figure \ref{fig:cheque_abm_t_100000} and \ref{fig:cheque_sme_t_100000}), the higher cross-adhesion between the two populations leads to the preferential attachment of one cell type to the other indirectly repelling cells of the same type leading to the energetically favourable state resembling a checkerboard.

Our model has the potential to elucidate cell adhesion dynamics in a two-species environment, applicable across a spectrum of biological contexts where cell-cell adhesion plays a role in driving cell cohesiveness such as wound healing, angiogenesis and the separation of cells in germ layers endoderm, mesoderm and ectoderm in the embryo \citep{dickinson2020tmc,schotz2008qdt}. 

Although we have included realism in our models by incorporating cell-cell interactions such as adhesion, volume filling and cell-cell swapping, we are neglecting the fact that cell populations are not constant over time. A natural extension of this study could be to allow cell division and analyse its effects. This could be achieved using a multistage cell-cycle model for improved realism \citep{yates2017mrc,smith1973cc,weber2014qlv,mort2014fbc,howard1951nip}. Another limitation is the fixed size of the domain over which cell movement is modelled. In reality, the domain (representing a developing embryo or growing tissue, for example) might grow over time. Incorporating domain growth and cell division into our model would improve the applicability of our work and help us understand the role of adhesion in a wider variety of biological processes.

\appendix

\section{Two-dimensional partial differential equations (PDEs)}\label{appendix_A}

Let $\hat A_{ij}(t)$ be the continuous approximation of the mean density of species A at the site $(i,j)$ at time $t$ and let $\hat B_{ij}(t)$ be the be the equivalent for species B. Then the master equation for the continuous approximation of the mean occupancy at the site $(i,j)$ and time $t+\delta t$ where $\delta t$ is an infinitesimally small change in time is given by,

\begin{align}\label{eqn:two_species_master_A}
&\hat A_{ij} (t+\delta t) = A_{ij}+\frac{m_A}{4} \delta t (1-\hat A_{ij}-\hat B_{ij})\Big[ \hat A_{i-1,j}(1-p)^{\Sigma_{z \in Z_{i-1,j}} \hat A_z} (1-r)^{\Sigma_{z \in Z_{i-1,j}} \hat B_z}\nonumber \\
& \quad + \hat A_{i+1,j} (1-p)^{\Sigma_{z \in Z_{i+1,j}} \hat A_z} (1-r)^{\Sigma_{z \in Z_{i+1,j}} \hat B_z} + \hat A_{i,j-1} (1-p)^{\Sigma_{z \in Z_{i,j-1}} \hat A_z} (1-r)^{\Sigma_{z \in Z_{i,j-1}} \hat B_z} \nonumber \\
& \qquad \qquad \qquad \qquad \qquad \qquad \qquad \qquad \qquad  + \hat A_{i,j+1} (1-p)^{\Sigma_{z \in Z_{i,j+1}} \hat A_z} (1-r)^{\Sigma_{z \in Z_{i,j+1}} \hat B_z}\Big] \nonumber \\
& \quad - \frac{m_A}{4}\delta t \hat A_{ij}(1-p)^{\Sigma_{z \in Z_{ij}} \hat A_z}(1-r)^{\Sigma_{z \in Z_{ij}} \hat B_z}\Big[4-\hat A_{i-1,j}-\hat B_{i-1,j} -\hat A_{i+1,j}-\hat B_{i+1,j} \nonumber \\
& \quad \qquad \qquad \qquad \qquad \qquad \qquad \qquad \qquad \qquad \qquad  -\hat A_{i,j-1}-\hat B_{i,j-1} -\hat A_{i,j+1}-\hat B_{i,j+1}\Big] \nonumber \\
& \quad + \frac{(m_A+m_B)}{4}\delta t \rho B_{ij} (1-q)^{\Sigma_{y \in Y_{ij}} \hat B_z} (1-r)^{\Sigma_{y \in Y_{ij}} \hat A_z} \Big[\hat A_{i-1,j}(1-p)^{\Sigma_{z \in Z_{i-1,j}} \hat A_z}(1-r)^{\Sigma_{z \in Z_{i-1,j}} \hat B_z} \nonumber \\
& \quad  + \hat A_{i+1,j} (1-p)^{\Sigma_{z \in Z_{i+1,j}} \hat A_z}(1-r)^{\Sigma_{z \in Z_{i+1,j}} \hat B_z} + \hat A_{i,j-1}(1-p)^{\Sigma_{z \in Z_{i,j-1}} \hat A_z}(1-r)^{\Sigma_{z \in Z_{i,j-1}} \hat B_z} \nonumber \\
& \quad \qquad \qquad \qquad \qquad \qquad \qquad \qquad \qquad \qquad + \hat A_{i,j+1}(1-p)^{\Sigma_{z \in Z_{i,j+1}} \hat A_z}(1-r)^{\Sigma_{z \in Z_{i,j+1}} \hat B_z} \Big] \nonumber \\
& \quad - \frac{(m_A+m_B)}{4}\delta t \rho \hat A_{i,j}(1-p)^{\Sigma_{z \in Z_{ij}} \hat A_z} (1-r)^{\Sigma_{y \in Y_{ij}} \hat B_z} \Big[\hat B_{i-1,j}(1-q)^{\Sigma_{y \in Y_{i-1,j}} \hat B_z} (1-r)^{\Sigma_{z \in Z_{i-1,j}} \hat A_z}\nonumber \\
& \quad + \hat B_{i+1,j}(1-q)^{\Sigma_{y \in Y_{ij}} \hat B_z}(1-r)^{\Sigma_{z \in Z_{i+1,j}} \hat A_z} + \hat B_{i,j-1}(1-q)^{\Sigma_{y \in Y_{ij}} \hat B_z}(1-r)^{\Sigma_{z \in Z_{i,j-1}} \hat A_z}\nonumber \\
& \quad \qquad \qquad \qquad \qquad \qquad \qquad \qquad \qquad \qquad + \hat B_{i,j+1}(1-q)^{\Sigma_{y \in Y_{ij}} \hat B_z}(1-r)^{\Sigma_{z \in Z_{i,j+1}} \hat A_z}\Big],
\end{align}

\begin{align}\label{eqn:two_species_master_B}
&\hat B_{ij} (t+\delta t) = B_{ij}+\frac{m_B}{4} \delta t (1-\hat B_{ij}-\hat A_{ij})\Big[ \hat B_{i-1,j}(1-q)^{\Sigma_{z \in Z_{i-1,j}} \hat B_z} (1-r)^{\Sigma_{z \in Z_{i-1,j}} \hat A_z}\nonumber \\
& \quad + \hat B_{i+1,j} (1-q)^{\Sigma_{z \in Z_{i+1,j}} \hat B_z} (1-r)^{\Sigma_{z \in Z_{i+1,j}} \hat A_z} + \hat B_{i,j-1} (1-q)^{\Sigma_{z \in Z_{i,j-1}} \hat B_z} (1-r)^{\Sigma_{z \in Z_{i,j-1}} \hat A_z} \nonumber \\
& \qquad \qquad \qquad \qquad \qquad \qquad \qquad \qquad \qquad  + \hat B_{i,j+1} (1-q)^{\Sigma_{z \in Z_{i,j+1}} \hat B_z} (1-r)^{\Sigma_{z \in Z_{i,j+1}} \hat A_z}\Big] \nonumber \\
& \quad - \frac{m_B}{4}\delta t \hat B_{ij}(1-q)^{\Sigma_{z \in Z_{ij}} \hat B_z}(1-r)^{\Sigma_{z \in Z_{ij}} \hat A_z}\Big[4-\hat B_{i-1,j}-\hat A_{i-1,j} -\hat B_{i+1,j}-\hat A_{i+1,j} \nonumber \\
& \quad \qquad \qquad \qquad \qquad \qquad \qquad \qquad \qquad \qquad \qquad  -\hat B_{i,j-1}-\hat A_{i,j-1} -\hat B_{i,j+1}-\hat A_{i,j+1}\Big] \nonumber \\
& \quad + \frac{(m_B+m_A)}{4}\delta t \rho A_{ij} (1-p)^{\Sigma_{y \in Y_{ij}} \hat A_z} (1-r)^{\Sigma_{y \in Y_{ij}} \hat B_z} \Big[\hat B_{i-1,j}(1-q)^{\Sigma_{z \in Z_{i-1,j}} \hat B_z}(1-r)^{\Sigma_{z \in Z_{i-1,j}} \hat A_z} \nonumber \\
& \quad  + \hat B_{i+1,j} (1-q)^{\Sigma_{z \in Z_{i+1,j}} \hat B_z}(1-r)^{\Sigma_{z \in Z_{i+1,j}} \hat A_z} + \hat B_{i,j-1}(1-q)^{\Sigma_{z \in Z_{i,j-1}} \hat B_z}(1-r)^{\Sigma_{z \in Z_{i,j-1}} \hat A_z} \nonumber \\
& \quad \qquad \qquad \qquad \qquad \qquad \qquad \qquad \qquad \qquad + \hat B_{i,j+1}(1-q)^{\Sigma_{z \in Z_{i,j+1}} \hat B_z}(1-r)^{\Sigma_{z \in Z_{i,j+1}} \hat A_z} \Big] \nonumber \\
& \quad - \frac{(m_B+m_A)}{4}\delta t \rho \hat B_{i,j}(1-q)^{\Sigma_{z \in Z_{ij}} \hat B_z} (1-r)^{\Sigma_{y \in Y_{ij}} \hat A_z} \Big[\hat A_{i-1,j}(1-p)^{\Sigma_{y \in Y_{i-1,j}} \hat A_z} (1-r)^{\Sigma_{z \in Z_{i-1,j} \hat B_z}}\nonumber \\
& \quad + \hat A_{i+1,j}(1-p)^{\Sigma_{y \in Y_{ij} \hat A_z}}(1-r)^{\Sigma_{z \in Z_{i+1,j} \hat B_z}} + \hat A_{i,j-1}(1-p)^{\Sigma_{y \in Y_{ij}} \hat A_z}(1-r)^{\Sigma_{z \in Z_{i,j-1}} \hat B_z}\nonumber \\
& \quad \qquad \qquad \qquad \qquad \qquad \qquad \qquad \qquad \qquad + \hat A_{i,j+1}(1-p)^{\Sigma_{y \in Y_{ij}} \hat A_z}(1-r)^{\Sigma_{z \in Z_{i,j+1}} \hat B_z}\Big].
\end{align}

Taylor expanding the terms in the above equations around $i$ and taking the diffusive limit leads to the coupled PDEs,

\begin{align*}
& \D A t = \nabla \cdot \Bigg[D_A\Big(D_1^\text{2D}(A,B) \nabla B +D_2^\text{2D}(A,B)\nabla A \Big)
\\ 
& \quad + (D_A+D_B)\rho \Big(D_3^\text{2D}(A,B)\nabla A - D_4^\text{2D}(A,B)\nabla B \Big)\Bigg],
\end{align*}

\begin{align*}
& \D B t = \nabla \cdot \Bigg[D_B \Big(D_5^\text{2D}(A,B) \nabla A + D_6^\text{2D}(A,B) \nabla B \Big) \\ & \quad + (D_A+D_B)\rho \Big(D_7^\text{2D}(A,B) \nabla B - D_8^\text{2D}(A,B) \nabla A \Big)\Bigg],
\end{align*}

where $D_A$, $D_B$ are given by Equation \eqref{eqn:diff_constant_2D} and $D_1^\text{2D},...,D_8^\text{2D}$ by Equation set \eqref{eqn:two_species_diff_coeff} in the main text.

\section{Two-dimensional stochastic mean equations}\label{appendix_B}

The two-dimensional stochastic mean equations (SMEs) can be derived directly from the master equations \eqref{eqn:two_species_master_A} and \eqref{eqn:two_species_master_B} by taking away $A_{ij}$ from both sides, dividing both sides by $\delta t$ and taking the limit as $\delta t \to 0$. This leads to $L_x\times L_y$ ordinary differential equations (ODEs) (one for each lattice site $(i,j)$),

\begin{align}
&\frac{\mathrm{d} \hat A_{ij}(t)}{\mathrm{d}t} = \frac{m_A}{4}  (1-\hat A_{ij}-\hat B_{ij})\Big[ \hat A_{i-1,j}(1-p)^{\Sigma_{z \in Z_{i-1,j}} \hat A_z} (1-r)^{\Sigma_{z \in Z_{i-1,j}} \hat B_z}\nonumber \\
& \quad + \hat A_{i+1,j} (1-p)^{\Sigma_{z \in Z_{i+1,j}} \hat A_z} (1-r)^{\Sigma_{z \in Z_{i+1,j}} \hat B_z} + \hat A_{i,j-1} (1-p)^{\Sigma_{z \in Z_{i,j-1}} \hat A_z} (1-r)^{\Sigma_{z \in Z_{i,j-1}} \hat B_z} \nonumber \\
& \qquad \qquad \qquad \qquad \qquad \qquad \qquad \qquad \qquad  + \hat A_{i,j+1} (1-p)^{\Sigma_{z \in Z_{i,j+1}} \hat A_z} (1-r)^{\Sigma_{z \in Z_{i,j+1}} \hat B_z}\Big] \nonumber \\
& \quad - \frac{m_A}{4} \hat A_{ij}(1-p)^{\Sigma_{z \in Z_{ij}} \hat A_z}(1-r)^{\Sigma_{z \in Z_{ij}} \hat B_z}\Big[4-\hat A_{i-1,j}-\hat B_{i-1,j} -\hat A_{i+1,j}-\hat B_{i+1,j} \nonumber \\
& \quad \qquad \qquad \qquad \qquad \qquad \qquad \qquad \qquad \qquad \qquad  -\hat A_{i,j-1}-\hat B_{i,j-1} -\hat A_{i,j+1}-\hat B_{i,j+1}\Big] \nonumber \\
& \quad + \frac{(m_A+m_B)}{4} \rho B_{ij} (1-q)^{\Sigma_{y \in Y_{ij}} \hat B_z} (1-r)^{\Sigma_{y \in Y_{ij}} \hat A_z} \Big[\hat A_{i-1,j}(1-p)^{\Sigma_{z \in Z_{i-1,j}} \hat A_z}(1-r)^{\Sigma_{z \in Z_{i-1,j}} \hat B_z} \nonumber \\
& \quad  + \hat A_{i+1,j} (1-p)^{\Sigma_{z \in Z_{i+1,j}} \hat A_z}(1-r)^{\Sigma_{z \in Z_{i+1,j}} \hat B_z} + \hat A_{i,j-1}(1-p)^{\Sigma_{z \in Z_{i,j-1}} \hat A_z}(1-r)^{\Sigma_{z \in Z_{i,j-1}} \hat B_z} \nonumber \\
& \quad \qquad \qquad \qquad \qquad \qquad \qquad \qquad \qquad \qquad + \hat A_{i,j+1}(1-p)^{\Sigma_{z \in Z_{i,j+1}} \hat A_z}(1-r)^{\Sigma_{z \in Z_{i,j+1}} \hat B_z} \Big] \nonumber \\
& \quad - \frac{(m_A+m_B)}{4} \rho \hat A_{i,j}(1-p)^{\Sigma_{z \in Z_{ij}} \hat A_z} (1-r)^{\Sigma_{y \in Y_{ij}} \hat B_z} \Big[\hat B_{i-1,j}(1-q)^{\Sigma_{y \in Y_{i-1,j}} \hat B_z} (1-r)^{\Sigma_{z \in Z_{i-1,j}} \hat A_z}\nonumber \\
& \quad + \hat B_{i+1,j}(1-q)^{\Sigma_{y \in Y_{ij}} \hat B_z}(1-r)^{\Sigma_{z \in Z_{i+1,j}} \hat A_z} + \hat B_{i,j-1}(1-q)^{\Sigma_{y \in Y_{ij}} \hat B_z}(1-r)^{\Sigma_{z \in Z_{i,j-1}} \hat A_z}\nonumber \\
& \quad \qquad \qquad \qquad \qquad \qquad \qquad \qquad \qquad \qquad + \hat B_{i,j+1}(1-q)^{\Sigma_{y \in Y_{ij}} \hat B_z}(1-r)^{\Sigma_{z \in Z_{i,j+1}} \hat A_z}\Big],
\end{align}

\begin{align}
& \frac{\mathrm{d}\hat B_{ij}(t)}{\mathrm{d}t}= \frac{m_B}{4}  (1-\hat B_{ij}-\hat A_{ij})\Big[ \hat B_{i-1,j}(1-q)^{\Sigma_{z \in Z_{i-1,j}} \hat B_z} (1-r)^{\Sigma_{z \in Z_{i-1,j}} \hat A_z}\nonumber \\
& \quad + \hat B_{i+1,j} (1-q)^{\Sigma_{z \in Z_{i+1,j}} \hat B_z} (1-r)^{\Sigma_{z \in Z_{i+1,j}} \hat A_z} + \hat B_{i,j-1} (1-q)^{\Sigma_{z \in Z_{i,j-1}} \hat B_z} (1-r)^{\Sigma_{z \in Z_{i,j-1}} \hat A_z} \nonumber \\
& \qquad \qquad \qquad \qquad \qquad \qquad \qquad \qquad \qquad  + \hat B_{i,j+1} (1-q)^{\Sigma_{z \in Z_{i,j+1}} \hat B_z} (1-r)^{\Sigma_{z \in Z_{i,j+1}} \hat A_z}\Big] \nonumber \\
& \quad - \frac{m_B}{4} \hat B_{ij}(1-q)^{\Sigma_{z \in Z_{ij}} \hat B_z}(1-r)^{\Sigma_{z \in Z_{ij}} \hat A_z}\Big[4-\hat B_{i-1,j}-\hat A_{i-1,j} -\hat B_{i+1,j}-\hat A_{i+1,j} \nonumber \\
& \quad \qquad \qquad \qquad \qquad \qquad \qquad \qquad \qquad \qquad \qquad  -\hat B_{i,j-1}-\hat A_{i,j-1} -\hat B_{i,j+1}-\hat A_{i,j+1}\Big] \nonumber \\
& \quad + \frac{(m_B+m_A)}{4} \rho A_{ij} (1-p)^{\Sigma_{y \in Y_{ij}} \hat A_z} (1-r)^{\Sigma_{y \in Y_{ij}} \hat B_z} \Big[\hat B_{i-1,j}(1-q)^{\Sigma_{z \in Z_{i-1,j}} \hat B_z}(1-r)^{\Sigma_{z \in Z_{i-1,j}} \hat A_z} \nonumber \\
& \quad  + \hat B_{i+1,j} (1-q)^{\Sigma_{z \in Z_{i+1,j}} \hat B_z}(1-r)^{\Sigma_{z \in Z_{i+1,j}} \hat A_z} + \hat B_{i,j-1}(1-q)^{\Sigma_{z \in Z_{i,j-1}} \hat B_z}(1-r)^{\Sigma_{z \in Z_{i,j-1}} \hat A_z} \nonumber \\
& \quad \qquad \qquad \qquad \qquad \qquad \qquad \qquad \qquad \qquad + \hat B_{i,j+1}(1-q)^{\Sigma_{z \in Z_{i,j+1}} \hat B_z}(1-r)^{\Sigma_{z \in Z_{i,j+1}} \hat A_z} \Big] \nonumber \\
& \quad - \frac{(m_B+m_A)}{4} \rho \hat B_{i,j}(1-q)^{\Sigma_{z \in Z_{ij}} \hat B_z} (1-r)^{\Sigma_{y \in Y_{ij}} \hat A_z} \Big[\hat A_{i-1,j}(1-p)^{\Sigma_{y \in Y_{i-1,j}} \hat A_z} (1-r)^{\Sigma_{z \in Z_{i-1,j} \hat B_z}}\nonumber \\
& \quad + \hat A_{i+1,j}(1-p)^{\Sigma_{y \in Y_{ij} \hat A_z}}(1-r)^{\Sigma_{z \in Z_{i+1,j} \hat B_z}} + \hat A_{i,j-1}(1-p)^{\Sigma_{y \in Y_{ij}} \hat A_z}(1-r)^{\Sigma_{z \in Z_{i,j-1}} \hat B_z}\nonumber \\
& \quad \qquad \qquad \qquad \qquad \qquad \qquad \qquad \qquad \qquad + \hat A_{i,j+1}(1-p)^{\Sigma_{y \in Y_{ij}} \hat A_z}(1-r)^{\Sigma_{z \in Z_{i,j+1}} \hat B_z}\Big].
\end{align}

\bibliographystyle{apsrev4-1}
\bibliography{/Users/srn32/Dropbox/PhD/All_my_papers/my_Library.bib}

\end{document}